\documentclass[12pt]{article}
\usepackage{amsmath, amssymb}  
\usepackage[paper=a4paper,%landscape,%
            lmargin=3.2cm,rmargin=3.0cm,tmargin=3.2cm,bmargin=3.2cm,%
            foot=0.9truecm]{geometry}
\def\mb#1{{\mathbf{#1}}}
\def\diag{\,\mbox{diag}\,}
\def\myot{\!\otimes\!}

\def\k{\kappa}

\def\so{{\mathfrak{so}}}
\def\iso{{\mathfrak{iso}}}
\def\su{{\mathfrak{su}}}
\def\u{{\mathfrak{u}}}
\def\sp{{\mathfrak{sp}}}
\def\sq{{\mathfrak{sq}}}
\def\usp{{\mathfrak{usp}}}

\def\sl{{\mathfrak{sl}}}
\def\sh{{\mathfrak{sh}}} % ???
\def\msh{{\mathfrak{msh}}}

\def\sa{{\mathfrak{sa}}}
\def\a{{\mathfrak{a}}} 

\def\Dos{2}

\def\esig{\cdot\cdot}
\def\sonl{\so_n^{\esig l \esig}}%   		so n l
\def\sodndl{\so_{2n}^{\esig 2l \esig}}%	so 2n 2l
\def\socncl{\so_{4n}^{\esig 4l \esig}}%	so 4n 4l
\def\sodnn{\so_{2n}^n}%   			so 2n n
\def\socndn{\so_{4n}^{2n}}%   			so 4n 2n
\def\sodnstar{\so_{2n}^*}%   			so 2n *
\def\socnstar{\so_{4n}^*}%   			so 4n *
\def\sonC{\so_n\Ce}%   				so n \C
\def\sodnC{\so_{2n}\Ce}%   			so 2n \C

\def\sunl{\su_n^{\esig l \esig}}%		su n l
\def\sudndl{\su_{2n}^{\esig 2l \esig}}%	su 2n 2l
\def\sudnn{\su_{2n}^n}%				su 2n n
\def\sudnstar{\su_{2n}^*}%			su 2n *
\def\slnR{\sl_n\Re}%				sl n \R
\def\sldnR{\sl_{2n}\Re}%				sl 2n \R
\def\slnC{\sl_n\Ce}%				sl n \C

\def\sqnl{\sq_n^{\esig l \esig}}%		sq n l
\def\sqdndl{\sq_{2n}^{\esig 2l \esig}}%	sq 2n 2l
\def\sqdnn{\sq_{2n}^n}%				sq 2n n
\def\spdnR{\sp_{2n}\Re}%				sp 2n \R
\def\spcnR{\sp_{4n}\Re}%				sp 4n \R
\def\spdnC{\sp_{2n}\Ce}%				sp 2n \C

\def\e{{\mathfrak{e}}}
\def\f{{\mathfrak{f}}}
\def\g{{\mathfrak{g}}}

\def\giic{\g_2^{(-14)}}
\def\giis{\g_2^{(2)}}
\def\fivc{\f_4^{(-52)}}
\def\fivxx{\f_4^{(-20)}}
\def\fivs{\f_4^{(4)}}
\def\evic{\e_6^{(-78)}}
\def\evixxvi{\e_6^{(-26)}}
\def\evixiv{\e_6^{(-14)}}
\def\eviII{\e_6^{(2)}}
\def\evis{\e_6^{(6)}}
\def\eviic{\e_7^{(-133)}}
\def\eviixxv{\e_7^{(-25)}}
\def\eviiv{\e_7^{(-5)}}
\def\eviis{\e_7^{(7)}}
\def\eviiic{\e_8^{(-248)}}
\def\eviiixxiv{\e_8^{(-24)}}
\def\eviiis{\e_8^{(8)}}

\def\hf{\,\hfil\,}
\def\ds{\,\dot\oplus\,}

\def\der{{\mathfrak{der}}\,}

\DeclareMathOperator{\RePart}{Re} %  \def\RePart{Re\hskip0.4pt}
\DeclareMathOperator{\ImPart}{Im} % \def\ImPart{Im\hskip0.4pt}

\def\Ke{{\mathbb K}} \def\K{{\mathbb K}}  %% Preferible usar la forma \Ke
\def\Ae{{\mathbb A}}

\def\Re{{\mathbb R}}  %% Tambien definido como \R
\def\Ce{{\mathbb C}}  %% Tambien definido como \C
\def\He{{\mathbb H}}  %% Tambien definido como \Q
\def\Oe{{\mathbb O}}  %% Tambien definido como \O
\def\CeS{{\mathbb C'}}   %% Tambien definido como \D
\def\HeS{{\mathbb H'}}   %% Tambien definido como \I
\def\OeS{{\mathbb O'}}   %% Tambien definido como \P

\def\Cei{\overline{\Ce}}
\def\CeSi{\overline{\CeS}}
\def\Her{\widetilde{\,\He\,}\!}
\def\HeSr{\widetilde{\HeS}}
\def\HeSh{\widehat{\HeS}}

\def\otimesc{\!\otimes\!} %%% old form \def\myot{\!\otimes\!}
\def\ve{\mathbf e}
\def\vx{\mathbf x}
\def\vy{\mathbf y}
\def\vz{\mathbf z}
\def\ovx{\overline{\mathbf x}}
\def\ovy{\overline{\mathbf y}}
\def\HermProd#1#2{\langle #1, #2 \rangle}
\def\CalA{{\cal{A}}}
\def\tr{\hbox{\rm tr}}

\def\matrixuno{\vuno}
\def\vuno{\mathbf 1}

\def\dichar#1#2#3{\lower-8pt\hbox{ #1}\ \hfill{ #2}\ \hfill {}^{(#3)}}

\def\vvrule{\vrule\hskip1pt\vrule}
\def\vruled{\vrule width 0.5pt}  % era 0.05pt no se porque
\def\vrulem{\vrule width 0.6pt}
\def\hruled{\hrule height 0.5pt}  % era 0.05pt no se porque
\def\hrulem{\hrule height 0.6pt}

\def\alg#1{{}_{#1}\Re}
\def\plus{\{+\}}
\def\minus{\{-\}}
\def\any{\{\!\!\{\pm\}\!\!\}}
\def\n{n}
\def\dn{2n}
\def\cn{4n}
\def\indexl{\{\!\!\{\n;l\}\!\!\}}
\def\indexdl{\{\!\!\{\dn;2l\}\!\!\}}
\def\indexcl{\{\!\!\{\cn;4l\}\!\!\}}
\def\indexN{\{\!\!\{\dn;\n\}\!\!\}}
\def\indexdN{\{\!\!\{\cn;\dn\}\!\!\}}
\def\indexti{\{\!\!\{3;1\}\!\!\}}
\def\SeparaBloque{\\[-8pt]\hline\\[-8pt]}
\def\mysa#1#2#3{#1 \, {\su}_{#2}(#3)}  %%% REEMPLAZA A LA DEF EN LecturaFicherosListados.tex
\def\mya#1#2#3{#1 \, {\u}_{#2}(#3)}    %%% REEMPLAZA A LA DEF EN LecturaFicherosListados.tex
\def\myref#1{(\ref{#1})}

\begin{document}

\markboth{M Santander}
{A perspective on the Magic Square}

%\title{A perspective on the Magic Square and the `special unitary' realizations of simple Lie algebras\footnote{Minor revision of the paper published in {\itshape Proceedings of the XXI International Fall Workshop on Geometry and Physics}, 	Int.\ J.\ Geom.\ Methods Mod.\ Phys., {\bf 10} 1360002, (September 2013), doi.org/10.1142/S0219887813600025}} %, World Scientific}}
%
%% Volume 10, Issue 08 (September 2013)
%% 1360002  https://doi.org/10.1142/S0219887813600025
%
%\author{ 
%Mariano Santander \\ [12pt]
%   {\itshape Departamento de F\'{\i}sica Te\'{o}rica and IMUVa, Facultad de Ciencias} \\
%   {\itshape Universidad de Valladolid, 47011 Valladolid, Spain} \\
%   {\itshape email: mariano.santander@uva.es}
%}
%%\date{August 30, 2022}
%-------------------------------------------------------------------
%\maketitle 

\centerline{\Large{A perspective on the Magic Square}}
\medskip

\centerline{\Large{and the `special unitary' realizations}}
\medskip

\centerline{\Large{of simple Lie algebras\footnote{Minor revision of the paper published in {\itshape Proceedings of the XXI International Fall Workshop on Geometry and Physics}, 	Int.\ J.\ Geom.\ Methods Mod.\ Phys., {\bf 10}, 08  (September 2013) 1360002, doi.org/10.1142/S0219887813600025}}} %, World Scientific}}

\bigskip\bigskip

% Volume 10, Issue 08 (September 2013)
% 1360002  https://doi.org/10.1142/S0219887813600025

%\author{ 
\centerline{\large{Mariano Santander}}

\bigskip

\centerline{{\itshape Departamento de F\'{\i}sica Te\'{o}rica and IMUVa, Facultad de Ciencias}}
\centerline{{\itshape Universidad de Valladolid, 47011 Valladolid, Spain}}
\centerline{{\itshape email: mariano.santander@uva.es}}
%}
%\date{August 30, 2022}
%-------------------------------------------------------------------
%\maketitle 

%%%%%%%%%%%%%%%%%%%%%%%%%%%%%%%%%%%%%%%%%%%%%%%%%%%%%%%%%%%%%%%%%%%%

\bigskip\bigskip

\begin{abstract}

This article contains the last part of the mini-course `Spaces: a perspective view' delivered at the IFWGP2012. The series of three lectures was intended to bring the listeners from the more naive and elementary idea of space as `{\itshape our physical Space}' (which after all was the dominant one up to the 1820s) through the  generalization of the idea of space which took place in the last third of the XIX century. That was a consequence of first the discovery and acceptance of non-Euclidean geometry and second, of the views afforded by the works of Riemann and Klein and continued since then by many others, outstandingly Lie and Cartan. 

Here I deal with the part of the mini-course which centers on the classification questions associated to the simple real Lie groups. I review the original introduction of the Magic Square `a la Freudenthal', putting the emphasis in the role played in this construction by the four normed division algebras $\Re, \Ce, \He, \Oe$. I then explore the possibility of understanding some simple real Lie algebras as `special unitary' over some algebras $\Ke$ or tensor products $\K_1\otimesc\K_2$, and I argue that the proper setting for this construction is not to confine only to normed division algebras, but to allow the split versions $\CeS, \HeS, \OeS$  of complex, quaternions and octonions as well. This way we get a `Grand Magic Square' and we fill in the details required to cover all real forms of simple real Lie algebras within this scheme. The paper ends with the complete lists of all realizations of simple real Lie algebras as `special unitary' (or only `unitary' when $n=2$) over some tensor product of two $*$-algebras $\K_1, \K_2$, which in all cases are obtained from $\Re, \Ce, \CeS, \He, \HeS, \Oe,\OeS$ as sets, endowing them with a $*$-conjugation which usually but not always is the natural complex, quaternionic or octonionic conjugation. 
\end{abstract}

%\keywords{Simple Lie algebras, magic squares, symmetric spaces, alternative algebras, composition algebras, Cayley-Klein geometry, Cayley-Dickson process}  %% xxx

\newpage

% >>>>>>>>>>>>>>>>>>>>>>>>>>>>>>>>>>>>>>>>>>>>>>>>>>>>>>>>>>>>>>>>>>

\section{Introduction}	

`Space' is probably one of the most overburdened terms nowadays used in Mathematics and in Physics. From a viewpoint which considers the long-time evolution of ideas, this is a comparatively rather recent phenomenon. When I got the invitation from the organizers of the IFWGP2012 to deliver a mini-course there, an invitation which I duly acknowledge, my idea was to provide a perspective view of this {\itshape `space to spaces'} development, discussing several modern aspects of the {\itshape `spaces'} idea, which are the closest to the historical meaning of the {\itshape Space} term. I planned to start from a swift historical presentation, covering from the prehistory of the space idea until the groundbreaking work by Riemann and Klein which actually set the starting point for the modern intelligence of the spaces. Then I planned to center attention in symmetric spaces, which provide a quite natural extension to the basic and deep properties that {\em Space} was assumed to have during its previous history. Surprisingly, (or perhaps not so much) these properties have also turned out to be essential in Physics. The following is a list of the topics which were discussed in the course of the three lectures, all through  interspersed with examples and comments on the ubiquitous presence of symmetric spaces all-around in Physics.  

\begin{itemize}

\item
The prehistory of the subject. 

\item
Basics on symmetric spaces: metric and connection. 

\item
Rank of a symmetric space. Beyond rank-one: Pl\"ucker geometry and phase spaces as rank-two spaces. 

\item
Classification of (irreducible, with simple isometry group) symmetric spaces as a follow up of the classification of Cartan list of simple real forms of Lie algebras. Understanding the Cartan and Berger lists: the generic real, complex and quaternionic spaces and the exceptional octonionic spaces. 

\item
The Magic Square of Lie algebras and the associated exceptional spaces: possible approaches.
\end{itemize}

For the written version I have chosen to deal only with the latter part of the actual talks, because the previous parts are well covered in many places. (For a general but detailed enough discussion, see \cite{HOS:TrigoSO, OS:TrigoSU} and specially \cite{HS:HomogeneousPhaseSpaces}; for some complementary information on several of the Cayley-Klein families one can refer to \cite{AHPS:CohomSO,HPS:CohomSU,HS:CohomSP} relative to  central extensions, \cite{HS:CasimirsSO} relative to the Casimirs, \cite{HMOS:GradedContSO,HS:GradedContSO} relative to graded contractions and also \cite{BHOS:Quantum2dCK,BHOS:Quantum3dCK} dealing with the associated quantum groups). I want to offer here a relatively self contained view of what could be called the `Cayley-Klein-Dickson' approach to the description of the real forms of simple Lie algebras and of their associated symmetric spaces. 
Even within this restriction, the text will be mainly descriptive, as going to the next level by including proofs would exceed the space reasonably allotted to the mini-course in these Proceedings; I hope to deal with all the detailed proofs in a separate extended version of this work. Some previous development appeared in \cite{HS:CK_FreudMagicSquares, S:SymHomSpacesViewpointClas, S:MatrixModels}.

% >>>>>>>>>>>>>>>>>>>>>>>>>>>>>>>>>>>>>>>>>>>>>>>>>>>>>>>>>>>>>>>>>>

\section{Symmetric spaces before Cartan}

The geometry of actual physical space had been explored and studied since antiquity, from many viewpoints and mixing its `physics' and `mathematics' sides, leading to its later formalization in the mathematical  concept of Euclidean space. The non-Euclidean epistemological breakthrough starting in the 1830s and spanning to the 1860s left a much clearer view  of the situation. If a few points are to be taken as the salient ideas, at least these two should be included: (i) {\em the Riemann viewpoint}: there are many `quadratic' geometries, not necessarily neither homogeneous nor isotropic, and the choice for the `physical' one should be an experimental matter; and  (ii) {\em the Klein viewpoint}: if we assume suitable forms of homogeneity and isotropy as restrictions, there are still many possibilities, but these are described completely through their `isometry groups'. 

Within the first viewpoint, a geometry is fully codified through a metric tensor. Within the second,  a geometry is fully codified by a group. Of course, the geometries of some special metric tensors with enough symmetries could be described also completely by their isometry group. All since its inception, there has been a tension among these two viewpoints, and both have proved to be essential for Physics: special relativity fits completely within the Klein viewpoint, with a Minkowskian geometry for space-time, while general relativity resorts to the Riemann viewpoint. These are not the only examples: quantum mechanics also fits within the Klein approach, as it is based in a complex unitary group as the invariance group. 

The connection between both approaches is a kind of `tangent approximation' (as embodied in the idea that  the actual (flexible) space-time of general relativity should be approximated at a tangent level (locally) by the (rigid) Minkowskian space-time of special relativity). Hence, while the study of geometries and spaces in the sense of Klein can be done for its own sake, their relevance as models for other geometries, where groups could quit the scene and only some more `flexible' geometry remains, should not be forgotten. 

The Euclidean geometry of the classical physical space, as well as the Minkowskian geometry behind special relativity are two {\itshape symmetric geometries}, and this is because there are some discrete transformations which turn out to be absolutely essential in their structures. For Euclidean geometry, these are reflections in a point, in a line and in a plane; for Minkowskian geometry we have space parity and time reversal. The Dieudonne-DeWitt theorem states that all isometries in these
geometries are generated by reflections, so it is sensible to put these reflections at the forefront. 
This leads to the Cartan definition of {\itshape symmetric space}: a homogeneous space $G/H$ of a Lie group $G$, where the subgroup $H$ is not an arbitrary subgroup of $G$ but must be the invariant subgroup under an {\itshape involutive automorphism} $\Theta$ of $G$. Euclidean geometry of physical space and Minkowskian geometry of space-time both  fit inside this scheme. 

There are three essentially equivalent characterizations.  There is a `Riemannian' definition: a symmetric space is a Riemannian (or pseudoRiemannian) space 
whose Riemann curvature tensor is covariantly constant. There is a direct geometric definition: a symmetric space is a Riemannian (or pseudoRiemannian) space  where the geodesic reflection around each point is an isometry; this allows a close link to the  description based on Lie groups, where the space is a coset space $G/H$.  Finally there is an  `algebraic' tangent definition, stated in terms of the Lie algebras $\mathfrak{g}$ and $\mathfrak{h}$ of $G$ and $H$: a symmetric space is a homogeneous space $G/H$ of a Lie group where at the Lie algebra level, the subalgebra $\mathfrak{h}$ must be the invariant subspace under some {\it involutive automorphism} of the algebra $\mathfrak{g}$.  

The `purely Riemannian' definition of a symmetric space provides the most obvious examples: the Euclidean spaces (flat Riemannian spaces with vanishing curvature) and the `Minkowskian' spaces (flat pseudoRiemannian spaces). These are symmetric spaces, and their isometry groups, in an $n$-dimensional ($n$-D) case,  are ${\rm ISO}(n), {\rm  ISO}(n-1,1)$ which are not simple Lie groups, but can be obtained through a well defined {\itshape contraction} procedure from the simple Lie groups ${\rm SO}(n+1), {\rm SO}(n,1)$ or ${\rm SO}(n, 1), {\rm SO}(n-1, 2)$, respectively. In particular, ${\rm SO}(n+1), {\rm SO}(n,1)$ are the isometry groups of the Riemannian spaces of non-zero constant curvature: the $n$-D sphere $\mb S^n \equiv {\rm SO}(n+1)/{\rm SO}(n)$, with constant sectional positive curvature and the hyperbolic space $\mb H^n\equiv {\rm SO}(n,1)/{\rm SO}(n)$, with constant sectional negative curvature.  The two groups ${\rm SO}(n, 1), {\rm SO}(n-1, 2)$ are the groups of isometries of the pseudoRiemannian locally Lorentzian spaces of constant curvature: the anti-deSitter ${\mb{AdS}}^{1\!+\!n}$  and deSitter ${\mb{dS}}^{1\!+\!n}$ spaces, with constant sectional positive curvature and negative curvature and a locally Lorentzian metric of signature type $(n,1)$.

All these symmetric spaces find a joint description within the {\itshape Cayley-Klein (CK) scheme}, where all these groups (the simple ones as well as their contractions), appear in a family of Lie groups (or algebras) parametrized by a CK list of labels ${\k_1, \k_2, \dots, \k_{n}}$. 
In \cite{HMOS:GradedContSO} one can find a complete description of this language, which deals at once with a full family of Lie algebras denoted $\so_{\k_1\k_2\cdots\k_{n}\!}(n+1)$. When all the CK labels are different from zero, this Lie algebra is simple, so isomorphic to some $\so_{n+1}^l$, and the negative inertia index $l$ is determined by the list of $\k_i$. 

The advantage of this language is that it puts to the forefront a set ot $n$ commuting involutions of the Lie algebra $\so_{\k_1\k_2\cdots\k_{n}\!}(n+1)$, so this description allows a direct construction of a set of $n$ symmetric spaces associated to the group ${\rm SO}_{\k_1\k_2\cdots\k_{n}\!}(n+1)$~\cite{IJTP}: for each involution $\theta_i$, the subalgebra $\mathfrak{h}$ can be immediately written and is given by $\so_{\k_2\cdots\k_{n}\!}(n)$ for $\theta_1$, by $\so_{\k_1}\oplus\so_{\k_3\cdots\k_{n}\!}(n-1)$ for $\theta_2$, etc. 

Their symmetric spaces which appear in this construction corresponding to the choice  of the involution $\theta_1$ are the $n$-D `spheres'  ${\rm SO}_{\k_1\k_2\cdots\k_{n}\!}(n+1)/{\rm SO}_{\k_2\cdots\k_{n}\!}(n)$ which for specific choices of the list $\k_i$ includes the usual sphere ${\rm SO}(n+1)/{\rm SO}(n)$ (for all $\k_i=1$), the hyperbolic space ${\rm SO}(n,1)/{\rm SO}(n)$ (for $\k_1=-1$, all other $\k_i=1$), the deSitter and anti-deSitter spheres, etc. The geometry and trigonometry of this family of spaces has been thoroughly studied in \cite{HOS:TrigoSO}, and has a physical bearing on the conformal symmetries and compactification of space-time, which has been discussed in \cite{HS:ConformalSymmetries} and  \cite{HS:ConformalCompact}. 

For the second involution $\theta_2$ we get all the Grassmannians of $1$-D lines in the previous space, which is ${\rm SO}_{\k_1\k_2\cdots\k_{n}\!}(n+1)/({\rm  SO}_{\k_1}(2)\otimes {\rm SO}_{\k_3\cdots\k_{n}\!}(n-1))$, etc. 

For the unitary complex analogous of the orthogonal Lie algebras, this approach has been developed in  \cite{OS:TrigoSU} and \cite{HPS:CohomSU} (see also \cite{OS:TrigoQSS}), and things are quite similar, with an interesting new fact: these belong to another CK family which has (for equal $n$) one more {\itshape label} $\eta_1$, called a Cayley-Dickson   label. When this label is positive, the algebras ${}_{\eta_1}\su_{\k_1\k_2\cdots\k_{n}\!}(n+1)$ are isomorphic to $\su_{n+1}^l$, and as each label (even the new one $\eta_1$) is associated to an involutive automorphism, by following the same construction we  get the complex projective elliptic and hyperbolic spaces, as well as all their pseudo-hermitian (with indefinite Hermitian metrics) versions and all the families of complex Grassmannians. But when the label $\eta_1$ is negative, the algebras ${}_{\eta_1}\su_{\k_1\k_2\cdots\k_{n}\!}(n+1)$ which appear in this case are isomorphic to $\sl_{n+1}\Re$. There is an  involutive automorphism ${}_1\theta$ of the algebra associated to this new label, and by applying to this automorphism the construction we discussed earlier we find a full family of symmetric spaces associated to this label: this family  includes e.g. ${\rm SU}(n+1)/{\rm SO}(n+1)$ (for  $\eta_1=1$ and all $\k_i=1$) and ${\rm SL}(n+1)/{\rm SO}(n+1)$ (for $\eta_1=-1$ and all $\k_i=1$). 

The family $\so_{\k_1\k_2\cdots\k_{n}\!}(n+1)$ can be seen as `special unitary' algebras over $\Re$ (orthogonal, in the ordinary parlance). The family ${}_{\eta_1}\su_{\k_1\k_2\cdots\k_{n}\!}(n+1)$ includes at once the  `special unitary' algebras over $\Ce$ (hence `special complex unitary') and the `special linear' algebras over the reals, which here appear as `special unitary' over {\itshape split complex numbers} $\CeS$ associated to the Cayley-Dickson label $\eta_1=-1$. In all these cases, a description of the algebras as `special unitary algebras' of (pseudo)-antihermitian matrices with entries in either $\Re$ or $\Ce, \CeS$ is the starting point to get a uniform description of most symmetric homogeneous spaces associated to these algebras. This brings us to the main question: can other real simple Lie algebras be realized in some similar form? This is the question we set.  

%_________________________________________________________
\subsection{Classification of simple real Lie algebras and associated symmetric spaces} 

When discussed at the Lie algebra level, the essential element characterizing the structure of a symmetric space is an involutive automorphism $\theta$ of a Lie algebra $\mathfrak{g}$: this will determine a subalgebra of $\mathfrak{g}$, denoted $\mathfrak{h}$, which is invariant under $\theta$, and through exponentiation one can build the corresponding Lie group $G$ and subgroup $H$ which provide the symmetric space as the coset space $G/H$. This corresponds to  the Cartan decomposition $\mathfrak{g} = \mathfrak{p} \oplus \mathfrak{h}$, with $\mathfrak{p}$ identifiable to the tangent space to $G/H$ at the point taken as origin, i.e., $H$ seen as a coset. 
 
A general classification of symmetric spaces is a quite difficult problem. This is further complicated by the fact that although there is a notion of {\itshape irreducibility} for symmetric spaces, there is no a  {\itshape canonical reduction} for reducible symmetric spaces. 

However, for simple Lie groups (whose symmetric spaces are {\it irreducible}), the classification  problem has been solved, and a complete list of the symmetric spaces associated to the compact simple Lie algebras was already given by Cartan, the list being completed with the spaces associated to non-compact simple Lie groups in the 1960s (the full list is in \cite{GilmoreBook}). The explicit and complete listing of  the irreducible symmetric homogeneous spaces associated to simple Lie groups depends on: 

\begin{itemize}
\item The classification of real simple Lie groups, hinted at by Killing and finally corrected and set out by Cartan \cite{Cartan14}, and 

\item The classification of involutive automorphisms of all simple Lie
algebras, listed by Cartan for compact algebras and completed by Gantmacher, Berger, Fedenko  and
others for the remaining non-compact algebras. 
\end{itemize}

As the Cartan classification of {\itshape real} simple Lie groups (or algebras) depends also on classifying the involutive automorphisms of the corresponding {\itshape complex} form, it turns out that the classification of symmetric spaces with real simple Lie group is somehow the {\itshape square} of the classification of {\itshape real simple Lie groups}. 

The Cartan classification of simple complex Lie algebras is summed up in the following list, which  includes all the simple Lie algebras over complex numbers (abbreviated simple complex Lie algebras; Cartan isomorphisms are not included): 
\begin{itemize}
\item Four infinite {\itshape classical} series $A_r, B_r, C_r, D_r$, $r=1, 2, \dots$

\item Five {\itshape exceptional} Lie algebras denoted as $G_2, F_4, E_6, E_7, E_8$.
\end{itemize}

In this standard Cartan notation, the capital letter denotes a family, usually called a Cartan series, and the subscript $r$ gives the rank of the algebra. 

Starting from the Cartan list of the simple complex Lie algebras, and by classifying its involutive automorphisms, one obtains the list of classification of simple real Lie algebras. That was also done completely by Cartan, and the result is that each complex Lie algebra has several possible {\itshape real forms}.  The complete list is:

\begin{itemize}

\item 
$A_{n-1}$: dimension {$n^2\!-\!1$}; split form $\sl_n\Re$, compact form $\su_n$; other real forms $\su_n^l, l=1, 2, \dots [n/2]$ and a further real form $\su_n^*$ when $n$ is even.  

\smallskip
\item 
$B_{n}$: dimension {$n(2n\!+\!1)$}; split form $\so_{2n+1}^n$, compact form
$\so_{2n+1}$; other real forms $\so_{2n+1}^l, l=1, 2, \dots n\!-\!1$.

\smallskip
\item 
$C_n$: dimension {$n(2n\!+\!1)$}; split form $\sp_{2n}\Re$, compact form
$\sq_n$, other real forms 
$\sq_n^l, l=1, 2, \dots [n/2]$.

\smallskip
\item 
$D_{n}$: dimension {$n(2n\!-\!1)$}; split form $\so_{2n}^n$, compact form
$\so_{2n}$; other real forms 
$\so_{2n}^l, l=1, 2, \dots n\!-\!1$ and $\so_{2n}^*$.

\smallskip
\item 
$G_2$: dimension {14};\ split form $\giis$, compact form $\giic$. 

\smallskip
\item 
$F_4$: dimension {52};\ split form $\fivs$, compact form $\fivc$, other
$\f_4^{(-20)}$.

\smallskip
\item 
$E_6$: dimension {78};\ split form $\evis$, compact form $\evic$, other $\e_6^{(-26)}$, $\e_6^{(-14)}$, $\e_6^{(2)}$.

\smallskip
\item 
$E_7$: dimension {133};\ split form $\eviis$, compact form $\eviic$, other $\e_7^{(-25)}, \e_7^{(-5)}$.

\smallskip
\item 
$E_8$: dimension {248};\  split form $\eviiis$, compact form $\eviiic$, other $\e_8^{(-24)}$.

\end{itemize}

Further to that, for each Cartan complex simple Lie group $G$, with complex dimension $d$, there is an associated real form which is $G$ seen as a real group; its real dimension is of course $2d$ and sometimes, due to its rather obvious nature, these groups are not even listed as real forms, although they are non-isomorphic to any of the previous ones. Going to their Lie algebras, these comprise  $\sl_n\Ce,\, \so_{2n+1}\Ce,\, \sp_{2n}\Ce$ and $\so_{2n}\Ce$ as well as the complex exceptional algebras seen as real Lie algebras.  

Some comments on this list and on the notation are called for. In the algebras $\so_n^l,\, \su_n^l,\, \sp_n^l$, the subscript $n$ refers to the dimension of the space $\Re^n, \, \Ce^n, \, \He^n$ where the corresponding group acts as (linear) isometries and $l$ is the inertia index;
%i.e., the number of diagonal terms with negative sign in the diagonal form of the $n\times n$ metric matrix; 
 the relation of this value with the standard physicist notation $\so(p,q), \, \su(p,q), \, \sq(p,q)$ is conveyed by $p\!+\!q=n,\, p\!>\!q,\, q\!=\!l$, so e.g., $\so_4^1\equiv \so(3,1)$. In the exceptional Lie algebras as $\e_7^{(-25)}$ the subscript gives the rank of the algebra and the superscript is the character of the real form (i.e., the signature of the Killing-Cartan metric which for the compact form equals to minus the algebra dimension, as the KC metric is negative definite).  
The `split' or `anticompact' name is given to the real form with the largest possible value of the character, which turns out to be equal to the rank. 

%_________________________________________________________
\subsection{`Understanding' the list of simple Lie algebras}

At the time when this classification was obtained some of the simple Lie algebras could be either related to already known geometric transformation groups (as e.g., $\so_n$) or to some spaces which were already studied at the time (as the complex unitary spaces behind $\su_n$, discussed by Fubini and Study). Other algebras  were related to these in some rather direct way (e.g., by allowing metrics with different signature but still non-degenerate, which amounts to allow all possible `inertia' indices), for instance $\so_n^l,\ \su_n^l$. There were also algebras related to ongoing geometric investigations, as the algebras of the initially so-called `complex groups', related to the geometry of a complex of lines; some time later this name was fortunately changed by Chevalley \cite{Chevalley}
to {\itshape symplectic groups} (see also \cite{Dieudonne}). Notation and naming for these are still a bit prone to confusion, as the term {\itshape symplectic} is also used for an antisymmetric scalar product, as related to the algebras $\sp_{2n}\Re,\ \sp_{2n}\Ce$ where it has a different meaning than in the `symplectic' $\sp_n^l$ or $\usp_n^l$, and this is the main reason to prefer a different notation $\sq_n^l$ for $\sp_n^l\equiv \usp_n^l$, as advocated by Sudbery \cite{Sudbery84}. 

The interesting point here is that the final complete list, further to the algebras of already known groups, included some previously unknown, completely new objects. Among them one finds a few new Lie algebras today classed as `classical' and, mainly, all exceptional algebras $\g_2, \f_4, \e_6, \e_7, \e_8$, each with several real forms. The list of real Lie algebras which at its discovery time had no any interpretation as linked to some groups of transformations were: 

\begin{itemize} 
\item 
$A_{2n-1}: \sudnstar$

\item 
$D_n: \sodnstar$

\smallskip

\item 
$\g_2:  \giic,\giis$

\item
$\f_4: \fivc,\fivxx,\fivs$

\item
$\e_6: \evic,\evixxvi,\evixiv,\eviII,\evis$

\item
$\e_7: \eviic, \eviixxv,\eviiv,\eviis$ 

\item
$\e_8: \eviiic,\eviiixxiv,\eviiis$
\end{itemize}

Understanding for these objects came slowly, and this can be expected, as these are worth of a place among the more complicated mathematical objects we know. In the next pages we will present a  scheme to get at least some place for them, following partially the historical development, which has been slow and complicated. 

%_________________________________________________________

\section{Cayley-Dickson doublings of the real numbers}

The standard Cayley-Dickson (CD) doubling process \cite{Dickson} can be applied to any $*$-algebra $\Ae$. In its `usual' form it adjoins a new unit $i$ to the algebra, which should satisfy the two basic sets of conditions
\begin{equation}
i^2 = -1, \qquad i^* = - i ,
\label{CDbasicConds}
\end{equation}
\begin{equation}
a (b i) = (b a) i, \qquad (a i) b = (a b^*) i, \qquad (a i) (b i) = -b^* a .
\label{CDaditConds}
\end{equation}
The conditions \myref{CDbasicConds} mean that $i$ must have square equal to $-1$ (such unit will be termed {\itshape elliptical}) and has to be pure imaginary in the doubled algebra (i.e., the $*$-conjugation acts as the old on $\Ae$, but in the new doubled algebra $i$ changes its sign under $*$). The conditions \myref{CDaditConds} mean that in the doubled algebra the $*$-conjugation  can be realized through ordinary `conjugation' by $i$. The three conditions in \myref{CDaditConds}  are to be separately required  when the doubled algebra is not associative; otherwise all three are equivalent \cite{Postnikov}.  
Elements in the doubled algebra can also be  seen as expressions as $a+ b i$ and are identified to pairs $(a, b) \in \Ae \times \Ae$, with  addition done componentwise, and multiplication and $*$-conjugation in the doubled algebra defined as
$$
i \equiv (0, 1), \qquad  (a, b) ( c, d) = (ac - d^*b, \ da + b c^*),\qquad  (a,b)^* = (a^*, -b) .
$$

The real numbers $\Re$ are trivially a $*$-algebra with the identity map as $*$. The complex numbers $\Ce$  are the first CD doubling of the real numbers $\Re$; this algebra is a {\itshape division, composition, $*$-algebra} which is {\itshape commutative} and {\itshape associative}. 

One can iterate the CD doubling: adjoin a new unit $j$ to $\Ce$ and require this unit $j$ to satisfy \myref{CDbasicConds} and \myref{CDaditConds} for all $a,b \in \Ce$. The result is the algebra of quaternions $\He$, discovered (along a different path) by Hamilton in 1843. This algebra, with real dimension $4$, is still a {\itshape division, composition, $*$-algebra}. Commutativity is however {\itshape lost} in favor of a specific form of non-commutativity: {\itshape pure imaginary quaternions are anticommutative}. Associativity is preserved. 

One can iterate again this process: adjoin a new unit $\ell$ to $\He$ and require this unit $\ell$ to be {\it elliptical} and pure imaginary. The result is the algebra of octonions $\Oe$, with real dimension $8$. This algebra is still a {\itshape division, composition, $*$-algebra.} Anticommutativity is preserved for pure imaginary octonions, but associativity is {\itshape lost} though not completely, and there is still some weak form of associativity: {\itshape octonions are only alternative.}
 
This process can be iterated again and again. This gives a full family of numeric systems, called $2^n$-onions. The next such stage provides the so-called {\itshape sedenions}, with real dimension $16$. Starting at sedenions the properties of being {\itshape a division and a composition algebra} are {\itshape lost} forever in the CD process. 

To sum up, when applied to the real numbers this process gives an infinite sequence of algebras of real dimensions $2, 4, 8, 16, \dots$, with the four first members being division algebras:  
\begin{equation}
\Re \to  \Ce \to  \He \to  \Oe \to  \hbox{Sedenions} \to   \dots 
\nonumber
\end{equation}
A deep result by Hurwitz states that $\Re, \Ce, \He, \Oe$ are the only normed division algebras \cite{Baez:Octonions}. 
As properties required in a normed division algebra seem to be important in our description of Nature, this mathematical result is also physically relevant. Hurwitz theorem  also suggests that there is some relevance in the CD process, seen as a constructive procedure which turns out to afford precisely the only four possible normed division algebras. 

%_________________________________________________________
\subsection{The `split' extension of the Cayley-Dickson doubling}

The standard CD process, as described above, is rigid, and it has been known since a long time that  we may introduce some `parametric freedom' in it. This gives some natural extensions which lead to different new algebras whose properties depart more and more from the division algebras as we introduce more and more changes in the construction. The first stage of this extension leads to some relatives of $\Ce, \He, \Oe$ which at first seem to be bizarre and unnatural extensions. However, their deep properties are so similar to the properties of $\Ce, \He, \Oe$ that actually can be seen as different instances of the same basic underlying structure.  
 
This most natural parametric freedom is to change the conditions \myref{CDbasicConds} to
$$
i^2 = \pm1, \qquad i^* = - i ,
\label{CDbasicCondsPar}
$$
where the adjoined unit must have a square which can be either equal to $-1$ or to $+1$, (in this last case the unit will be said {\itshape  hyperbolic unit}), while still being pure imaginary. Within this scheme, at each stage we have two possible choices instead of having just a single possibility. The first three stages of this extended process provide the result displayed in the scheme, where diagonal up (down) arrows correspond to addition of an elliptic (hyperbolic) unit

There are clearly two `brands' of complex numbers (ordinary complex $\Ce$, with $i^2=-1$ and {\itshape split complex} denoted $\CeS$, which are numbers of the form $a + i b $, with $i^2=1$ and the rest of the operatorial rules as complex numbers). The two `brands' of complex numbers are obviously different systems: for the split complex numbers one has $(1+i)(1-i) = 0$, hence $\CeS$ has divisors of $0$, which do not exist in $\Ce$. 

{\footnotesize
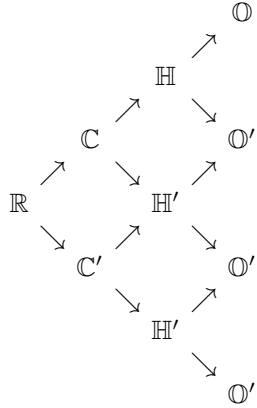
{
\begin{equation}
\begingroup
\arraycolsep 2pt
\begin{array}{cccccccc}
\ & \ &\ &\ &\ &\  &\ &\Oe\\
\ & \ &\ &\ &\ &\  &\nearrow &\\
\ & \ &\ &\ &\ &\He  &\ &\\
\ & \ &\ &\ & \nearrow &\  &\searrow &\\
\ & \ &\ &\Ce &\ &\  &\ &\OeS\\
\ & \ &\nearrow &\ &\searrow &\  &\nearrow &\ \\
\ & \Re &\ &\ &\ &\HeS &\ &\\
\ & \ &\searrow &\ &\nearrow &\  &\searrow &\ \\
\ & \ &\ &\CeS &\ &\  &\ &\OeS\\
\ & \ &\ &\ & \searrow &\  &\nearrow &\\
\ & \ &\ &\ &\ &\HeS  &\ &\\
\ & \ &\ &\ &\ &\  &\searrow &\\
\ & \ &\ &\ &\ &\  &\ &\OeS\\
\end{array}
\endgroup\nonumber
\end{equation}
}}

Naively one might perhaps expect three brands of quaternions and four brands of octonions, but actually there are only {\itshape two} brands of each. The reason is clear: in the quaternions, if $i$ and $j$ are elliptical, so is $k:=i j$; if only one of $i, j$ is hyperbolical, so is $k$, and finally if both $i, j$ are hyperbolical, $k$ is necessarily elliptical; thus the only possibilities  for different brands of quaternions are: (i) the {\itshape ordinary quaternions} $\He$ with a total of three elliptical units, and (ii) the {\itshape split quaternions}, $\HeS$ with one elliptical and two hyperbolic units. A similar situation happens for octonions, where split octonions have three elliptical and four hyperbolic units. 

Real and complex numbers do not need any comment. Split complex numbers are quite different from (ordinary) complex numbers in some (important) properties but quite similar in their arithmetic description. For instance, $\Ce$ is a division algebra, $\CeS$ is not, but both $\Ce, \CeS$ are composition algebras. Essentially, this is similar to the situation happening with compact and non-compact real forms of the same groups or algebras: some important properties are quite different, yet commutation rules encoding the Lie algebra structure are quite similar. This supports the view that $\Ce$ and $\CeS$ can be seen as two different instances of essentially the same structure, so from now on, generic mentions of `complex' not adequately qualified should be understood as referring to both `ordinary and split complex', with the same convention for quaternions and octonions.  

Quaternions $\He$ are also well known, and   their main difference to complex numbers lies in the anticommutativity of the three imaginary units. As quaternions are still associative, they are not so far from familiar land (after all, the Pauli matrices can be secretly seen as pure imaginary unit quaternions). 

Then we arrive to octonions $\Oe$ and to its split variant $\OeS$. What is new for octonions? Octonions $\Oe$  are the last CD doubling to be a division algebra (of course, $\OeS$ is not a division algebra). Both $\Oe, \OeS$ are composition algebras, as are both types of complex $\Ce, \CeS$ and of quaternions $\He, \HeS$. Octonions are the first `eccentric' member of the algebraic family of CD doublings, as they are not associative. But to insist in this lack of associativity is to tell only part of the truth, because octonions still possess some remnant of associativity, called {\itshape alternativity} and it is better to put this property to the forefront. 

%_________________________________________________________
\subsection{Basic properties of alternative composition algebras}

In any algebra $\Ae$, lack of commutativity or  of associativity can be  respectively measured by the {\itshape commutator} $[\vx, \vy]$ and {\itshape associator} $[\vx, \vy, \vz]$  defined for $\vx, \vy, \vz \in\Ae$ as
$$
[\vx, \vy] := \vx \vy - \vy \vx, \qquad
[\vx, \vy, \vz] := (\vx \vy) \vz - \vx (\vy \vz). 
\label{DefCommAss}
$$
Commutator is a bilinear map and associator is a trilinear map, which identically vanish respectively for commutative and associative algebras. 

For our algebras $\Re, \Ce, \CeS, \He, \HeS, \Oe, \OeS$, commutators and associators vanish when one of its arguments is real. The commutator is naturally alternating, i.e., it always changes sign in each swapping of any two arguments. {\itshape Alternative algebras} are defined as algebras with an {\itshape alternating} associator, i.e., an associator which changes sign in each swapping of any two arguments
$$
[\vx, \vy, \vz] = -[\vx, \vz, \vy], \qquad 
[\vx, \vy, \vz] = -[\vz, \vy, \vx], \qquad
[\vx, \vy, \vz] = -[\vy, \vx, \vz] .
\label{DefAltern}
$$

Although it may be not completely clear at first, it turns out that alternativity should be seen as some weak or remnant form of associativity. Why? The reason is that in alternative algebras some `restricted associativity' still exists in products of three octonions only two of which are different (Artin identities)
$$
(\vx \vx) \vy =\vx (\vx \vy), \quad 
(\vx \vy) \vx =\vx (\vy \vx), \quad
(\vy \vx) \vx =\vy (\vx \vx), 
\label{ArtinIds}
$$
and in products of four octonions only three of which are different  (Moufang identities)
$$
\big( \vx (\vy \vx) \big) \vz = \vx \big(\vy (\vx\vz) \big), \quad 
\vx (\vy \vz) \vx =(\vx \vy)(\vz \vx), \quad 
\vy\big((\vx\vz) \vx \big) = \big((\vy\vx)\vz \big)\vx .
\label{MoufangIds}
$$

Of course, any associative algebra is automatically (and trivially) alternative. Now the essential point is that, while octonions are actually not associative, insisting in this aspect might hide the fact that octonions still possess {\itshape alternativity}, which can be seen as the shadow of associativity after a stage in the CD process (to be precise, in the CD doubling, the doubling of $\Ae$ is alternative and nicely normed whenever $\Ae$ itself is (fully) associative and nicely normed;  see Prop. 4 in \cite{Baez:Octonions}). As octonions are not associative, the sedenions, which are the next stage in the CD process are not even alternative, so alternativity is lost after the fourth CD stage. 

Hence, the natural conditions which are satisfied precisely by the algebras in the set 
$
\Re \quad \begin{array}{l} \Ce \\ \CeS \end{array} \quad 
\begin{array}{l} \He \\ \HeS \end{array} \quad 
\begin{array}{l} \Oe \\ \OeS \end{array}
$
and only for these are to be {\itshape alternative, composition $*$-algebras}. From now on  these algebras will be denoted collectively by the symbol $\Ke$. Let us look to some properties of such systems. 

On any $\Ke = \Re, \Ce, \CeS, \He, \HeS, \Oe, \OeS$  there is an {\itshape inner product} (a symmetric real bilinear form in the underlying real vector space) which is given by 
$$
\HermProd{\vx}{\vx}:=\RePart(\vx \ovx)=\vx \ovx, \qquad 
\HermProd{\vx}{\vy}:=\RePart(\vx \ovy). 
$$
This inner product is {\itshape non-degenerate} for all our $\Ke$. It is {\itshape definite positive} for  $\Re, \Ce, \He, \Oe$, while  for the split versions $\CeS, \HeS, \OeS$ the inner product is indefinite, with signatures $(1,1)$, $(2,2)$ and $(4,4)$. 

Second, all these algebras $\Ke$   satisfy a {\itshape composition property} which  turns out to be very restrictive. In terms of the quadratic form associated to the bilinear inner product, the {\itshape composition algebra} condition requires that for any $\vx, \vy\in\Ke$  the following identity is satisfied: 
$$
\HermProd{\vx\vy}{\vx\vy}=\HermProd{\vx}{\vx} \HermProd{\vy}{\vy}. 
\label{CompCond}
$$

For complex numbers the composition reduces to the multiplicative property of the complex modulus, a property which is equivalent to the two squares identity known from the antiquity.
For quaternions, the composition property boils down to the four squares identity discovered by Euler. Finally, for octonions, the composition condition is equivalent to the eight squares identity first discovered by Degen and then rediscovered a quarter of century later by Graves and Cayley (see \cite{Curtis48Squares}).

The commutator of two elements and the associator of three elements in $\Ke$ can be equivalently seen as a family  of maps from $\Ke$ to $\Ke$: the {\itshape commutator maps} $C_\vx$ and the {\itshape associator maps} $\CalA_{\vx, \vy}$ which are defined as the linear applications from $\Ke$ to $\Ke$ given by
$$
C_\vx(\vz) := \vx \vz - \vz \vx, \qquad \CalA_{\vx, \vy}(\vz) := [\vx, \vy, \vz],
$$
and although these definitions may seem redundant, we will make an essential use of them. 
By direct computation, one may check that the maps ${\cal D}_{\vx, \vy}$
 defined by
\begin{equation}
{\cal D}_{\vx, \vy}=C_{[\vx, \vy]}-3 \CalA_{\vx, \vy},
\label{DerOctFromCA}
\end{equation}
are {\itshape derivations} of $\Ke$ for all our $\Ke$ (i.e., they satisfy Leibniz condition ${\cal D}(\vx \vy) = {\cal D}(\vx)\, \vy + \vx \,{\cal D}(\vy)$), and by direct brute-force calculation, it turns out that any derivation of $\Ke$ can be written a sum of derivations of the type ${\cal D}_{\vx, \vy}$. 

Now it is time for a quick overview stressing the features which will turn out to be relevant later: 

\begin{itemize}
\item
The algebras $\Re, \Ce, \He, \Oe$ and the split variants $\CeS, \HeS, \OeS$ are the only {\itshape composition, alternative $*$-algebras}. 

\item
Among them only $\Re, \Ce, \He, \Oe$ are {\itshape division algebras}, with an inner product which is positive definite. These algebras allow a {\itshape norm} in the usual sense, and for them all the pure imaginary units are elliptical. By Hurwitz theorem, these four algebras are the only {\itshape normed division algebras}.

\item
The split variants $\CeS, \HeS, \OeS$ have divisors of zero and therefore are not division algebras. For them the inner product is indefinite, and there are isotropic vectors. There is no a norm on these $\Ke$ in the usual sense, and for them at least one of the units adjoined in the stages of the CD process is hyperbolical, which means that in these cases $\Ke$ has both elliptic and hyperbolic pure imaginary units.    
\end{itemize}

As we shall see, the property of being a normed or division algebra is {\it not} the  essential token to some of the constructions we are going to discuss, and a clearer view can be obtained if we do not enforce the restriction to normed division algebras but allow all composition alternative $*$-algebras in the game. As commented earlier, the restriction from the $\Ke$ family to those algebras which have the division property is tantamount to the restriction to the compact form in a family of Lie algebras with the same complex form; even if there are fundamental distinctions in some properties, as far as the construction process itself, the differences are rather minor.  

% >>>>>>>>>>>>>>>>>>>>>>>>>>>>>>>>>>>>>>>>>>>>>>>>>>>>>>>>>>>>>>>>>>

\section{The historical approach to the `classical' Magic Square through unitary, linear and Hermitian symplectic Lie algebras}

To start with, the algebras $\so_n^l$ are related with the geometry of a real symmetric scalar product. Similarly, $\su_n^l$ is related to the geometry of a complex `Hermitian symmetric' scalar product. Both families of algebras include the geometries of a symmetric or Hermitian scalar product, corresponding to all possible signatures, as the associated quadratic or Hermitian form has inertia index $l$ ($n-l$ positive terms and $l$ negative ones when reduced to its diagonal form). 

A natural question at this point is: if these algebras are related to real and complex numbers, $\Re$ and $\Ce$, are there any algebras in a similar relation to the quaternions $\He$? 
By the 1950s, it was already fully clear that real and complex geometries had $\He$ relatives. The non-commutativity of $\He$ requires some care (one has to take a consistent procedure as to extract scalars from quaternionic vectors or multiply quaternionic vectors by scalars). Once adequately set, everything works fine also for geometries based on a `quaternion Hermitian' scalar product on $\He$-vector spaces, and the fact that quaternionic conjugation is an involutory {\itshape antiautomorphism} is essential. Complex conjugation is actually an involutory antiautomorphism of the complex numbers, seen simply as an automorphism due to commutativity of $\Ce$. There are of course completely new traits related to the existence of non-trivial automorphisms of quaternions, with no analogues for the real and complex cases. 

The end result of this analysis is that the family of algebras in the $C_n$ Cartan series, usually denoted $\usp_n^l$ could  (should?) be looked to precisely as the `special unitary algebras over quaternions'. To avoid confusion, it is better to rename these algebras abstractly as $\sq_n^l$, as proposed by Sudbery; its relationship to quaternions is stated in the identification $\He \su_n^{\esig l \esig} \equiv \sq_n^l$. 
When we look more closely to simple Lie algebras from this `real, complex(ification) and  quaternion(ization)' perspective, we get a vantage viewpoint from which one could reach a fascinating and intriguing object, the so-called `classical' Magic Square of Lie algebras. 

Let us arrange in a row the `special unitary' Lie algebras over $\Re, \Ce, \He$, this is, the algebras $\Re\su_n^l,\ 
\Ce\su_n^l,\ \He\su_n^l$. These are the Lie algebras of the isometries of $(n\!-\!1)$-D spherical geometry and its $\Ce$ and $\He$ Hermitian relatives. At each entry we will include some information on the Lie algebra, to be commented upon below. 

Each box-entry in the Table includes a lower line with the standard name of the algebra and its identification with a `special unitary' algebra over $\Re, \Ce$ or a `unitary' algebra over $\He$ (which is still `special' in a sense to be described below), in the middle line the Cartan series to which each Lie algebra belongs (the superscript ${\esig l \esig}$ refers to the possibility of different real forms, with different inertia indices, or when pertinent later, characters to be written in parentheses)  and in the left upper corner we include the real dimension of the Lie algebra. 

\bigskip
{%\footnotesize
\centerline{\vbox{\offinterlineskip
\def\tr{\omit&height 1pt&&&&&&} %%% Local \defs, working only inside the group 
\def\mr{\omit&height 6pt&&&&&&}
\def\myquad{\qquad}
\def\myqquad{\qquad}
\halign{
\strut$#$\quad&\vvrule#\ &\hf$#$\hf&\vrulem#\ &\hf$#$\hf&\vruled#\ &\hf$#$\hf&\vrulem#\ &\hf$#$\hf&\vrulem#\vrulem\cr
           & & \myquad\Re\myquad  & & \ \myquad\myquad\Ce\myquad\myquad\   & & \myquad\myquad\He\myquad\myquad   &  \cr
\noalign{\hrule\vskip1pt\hrule}
\mr\cr
    & & \dichar{$\tfrac{n(n-1)}{2}$}{D_{\tfrac{n}{2}} \ \hbox{or}\  B_{\tfrac{n-1}{2}}}{\esig } & & \dichar{$n^2\!-\!1$}{A_{n-1}}{\esig }  &&  \dichar{$n(2n\!+\!1)$}{C_n}{ \esig}   & \cr\tr\cr
\su    & & \rule[-10pt]{0.0pt}{30pt} \Re\su_n^{\esig l \esig} \equiv \sonl  & & \Ce\su_n^{\esig l \esig} \equiv\sunl     &&  \He\su_n^{\esig l \esig} \equiv\sqnl   &   \cr\mr\cr
\noalign{\hrulem}
%\mr\cr
%    & & \dichar{$n^2\!-\!1$}{A_{n-1}}{n-1}  & & \dichar{$2n^2\!-\!1$}{2A_{n-1}}{0}   &&  \dichar{$4n^2\!-\!1$}{A_{2n-1}}{-2n-1}   &   \cr\tr\cr
%\sl   & & \rule[-10pt]{0.0pt}{30pt} \R \sl_n\equiv\slnR & & \C \sl_n\equiv\slnC   && \He \sl_n\equiv\sudnstar &  \cr\mr\cr
%\noalign{\hruled}
%\mr\cr
%    & & \dichar{$n(2n\!+\!1)$}{C_n}{n} & & \dichar{$4n^2\!-\!1$}{A_{2n-1}}{1}  &&  \dichar{$2n(4n\!-\!1$)}{D_{2n}}{-2n}   &  \cr\tr\cr
%\sh   & & \rule[-10pt]{0.0pt}{30pt} \R \sh_{2n}\equiv\spdnR  & & \C \sh_{2n}\equiv\sudnn  &&  \He \sh_{2n}\equiv\socnstar &   \cr\mr\cr
\noalign{\hruled}
}}
}
}

\bigskip

\enlargethispage{\baselineskip}
Now let us add another line to this Table, labelled $\sl$ and which will include the Lie algebras of {\itshape special linear groups}. These are quite well known for vector spaces on $\Re$ and $\Ce$, and now acquire a new relative, the special linear group for a quaternionic vector space of dimension $n$, over $\He$. The analysis leads to the identification $\He\sl_n\equiv \sudnstar$, therefore providing an interpretation for one of the `rare' Lie algebras in the Cartan classification. 
In this row  there is no inertia index, and therefore only a single Lie algebra appears, whose character is given as a superscript in parentheses 
We now add this second row to the former scheme: 

\bigskip
{%\footnotesize
\centerline{\vbox{\offinterlineskip
\def\tr{\omit&height 1pt&&&&&&}%%% Local \defs, working only inside the group
\def\mr{\omit&height 6pt&&&&&&}
\def\myquad{\qquad}
\def\myqquad{\qquad}
\halign{
\strut$#$\quad&\vvrule#\ &\hf$#$\hf&\vrulem#\ &\hf$#$\hf&\vruled#\ &\hf$#$\hf&\vrulem#\ &\hf$#$\hf&\vrulem#\vrulem\cr
           & & \myquad\Re\myquad  & & \ \myquad\myquad\Ce\myquad\myquad\   & & \myquad\myquad\He\myquad\myquad   &  \cr
\noalign{\hrule\vskip1pt\hrule}
\mr\cr
    & & \dichar{$\tfrac{n(n-1)}{2}$}{D_{\tfrac{n}{2}} \ \hbox{or}\  B_{\tfrac{n-1}{2}}}{\esig } & & \dichar{$n^2\!-\!1$}{A_{n-1}}{\esig }  &&  \dichar{$n(2n\!+\!1)$}{C_n}{ \esig}   & \cr\tr\cr
\su    & & \rule[-10pt]{0.0pt}{30pt} \Re\su_n^{\esig l \esig} \equiv \sonl  & & \Ce\su_n^{\esig l \esig} \equiv\sunl     &&  \He\su_n^{\esig l \esig} \equiv\sqnl   &   \cr\mr\cr
\noalign{\hrulem}
\mr\cr
    & & \dichar{$n^2\!-\!1$}{A_{n-1}}{n-1}  & & \dichar{$2n^2\!-\!1$}{2A_{n-1}}{0}   &&  \dichar{$4n^2\!-\!1$}{A_{2n-1}}{-2n-1}   &   \cr\tr\cr
\sl   & & \rule[-10pt]{0.0pt}{30pt} \Re \sl_n\equiv\slnR & & \Ce \sl_n\equiv\slnC   && \He \sl_n\equiv\sudnstar &  \cr\mr\cr
\noalign{\hruled}
%\mr\cr
%    & & \dichar{$n(2n\!+\!1)$}{C_n}{n} & & \dichar{$4n^2\!-\!1$}{A_{2n-1}}{1}  &&  \dichar{$2n(4n\!-\!1$)}{D_{2n}}{-2n}   &  \cr\tr\cr
%\sh   & & \rule[-10pt]{0.0pt}{30pt} \R \sh_{2n}\equiv\spdnR  & & \C \sh_{2n}\equiv\sudnn  &&  \He \sh_{2n}\equiv\socnstar &   \cr\mr\cr
%\noalign{\hruled}
}}
}
}

\bigskip

The next stage is to  consider the geometries associated to a `scalar product' with {\itshape Hermitian antisymmetry}, over the reals, complex and quaternions. The corresponding Lie algebras will be denoted (by using a purposely chosen non-standard notation) as  $\Re\sh_{2n},\ \Ce\sh_{2n},\ \He\sh_{2n}$. It is clear that $\Re\sh_{2n}\equiv \sp_{2n}\Re$ because `Hermitian antisymmetry' over $\Re$ is simply antisymmetry; real symplectic geometry therefore enters this game. Again, one can consider the $\Ce, \He$ relatives which due to the antiautomorphism property of the complex or quaternionic conjugation can be properly defined also in the quaternion case. In the complex case it is not difficult to check that $\Ce\sh_{2n}\equiv\sudnn$ so the algebra $\Ce\sh_{2n}$ is indeed isomorphic to the split (inertia index $n$) unitary algebra in the family $\su_{2n}^l$. In the quaternionic case, a detailed analysis identifies the Lie algebra $\He \sh_{2n}$ as $\He \sh_{2n}\equiv\socnstar$, something which sheds some light on another of the `rare' Lie algebras in the Cartan classification.  

\bigskip
{%\footnotesize
\centerline{\vbox{\offinterlineskip
\def\tr{\omit&height 1pt&&&&&&}%%% Local \defs, working only inside the group
\def\mr{\omit&height 6pt&&&&&&}
\def\myquad{\qquad}
\def\myqquad{\qquad}
\halign{
\strut$#$\quad&\vvrule#\ &\hf$#$\hf&\vrulem#\ &\hf$#$\hf&\vruled#\ &\hf$#$\hf&\vrulem#\ &\hf$#$\hf&\vrulem#\vrulem\cr
           & & \myquad\Re\myquad  & & \ \myquad\myquad\Ce\myquad\myquad\   & & \myquad\myquad\He\myquad\myquad   &  \cr
\noalign{\hrule\vskip1pt\hrule}
\mr\cr
    & & \dichar{$\tfrac{n(n-1)}{2}$}{D_{\tfrac{n}{2}} \ \hbox{or}\  B_{\tfrac{n-1}{2}}}{\esig } & & \dichar{$n^2\!-\!1$}{A_{n-1}}{\esig }  &&  \dichar{$n(2n\!+\!1)$}{C_n}{ \esig}   & \cr\tr\cr
\su    & & \rule[-10pt]{0.0pt}{30pt} \Re\su_n^{\esig l \esig} \equiv \sonl  & & \Re\su_n^{\esig l \esig} \equiv\sunl     &&  \He su_n^{\esig l \esig} \equiv\sqnl   &   \cr\mr\cr
\noalign{\hrulem}
\mr\cr
    & & \dichar{$n^2\!-\!1$}{A_{n-1}}{n-1}  & & \dichar{$2n^2\!-\!1$}{2A_{n-1}}{0}   &&  \dichar{$4n^2\!-\!1$}{A_{2n-1}}{-2n-1}   &   \cr\tr\cr
\sl   & & \rule[-10pt]{0.0pt}{30pt} \Re \sl_n\equiv\slnR & & \Ce\sl_n\equiv\slnC   && \He \sl_n\equiv\sudnstar &  \cr\mr\cr
\noalign{\hruled}
\mr\cr
    & & \dichar{$n(2n\!+\!1)$}{C_n}{n} & & \dichar{$4n^2\!-\!1$}{A_{2n-1}}{1}  &&  \dichar{$2n(4n\!-\!1$)}{D_{2n}}{-2n}   &  \cr\tr\cr
\sh   & & \rule[-10pt]{0.0pt}{30pt} \Re \sh_{2n}\equiv\spdnR  & & \Ce\sh_{2n}\equiv\sudnn  &&  \He \sh_{2n}\equiv\socnstar &   \cr\mr\cr
\noalign{\hruled}
}}
}
}

\bigskip

Notice that neither the second nor the third row in this Table has `signature index' and in each box-entry there is a single Lie algebra, whose character is given as before in parentheses on the right side of the middle line. 

In this Table ---or rather we should say `tower of tables', as there is a table for each $n$--- we observe several intriguing properties: 

\begin{itemize}

\item
The Table is symmetric around the main diagonal as far as Cartan series are concerned (but not exactly as far as characters are concerned). 

\item
There are some curious numerical regularities here. For instance, the dimensions and the characters fit into some simple numerical schemes \cite{Freudenthal}. 
\end{itemize}

The Table makes sense for any $n=2, 3, \dots$, and it is not difficult to figure how to make sense out of it even for $n=1$, where the naive reading would say that some of the Lie algebras which appear there are zero-dimensional: indeed this is the only possible meaning to be ascribed to the symbols $\so_1$, $\su_1$, and $\sl_1\Re$; however the Lie algebras appearing in the $\He$ column and in the last row are non-trivial; e.g. $\He\su_1$ and $\He\sl_1$  are isomorphic  with dimension $3$. Much more of the structure which lies behind this glimpse will be told later. 

One may well ask whether these arrangements for some Lie algebras are a simple curiosity which arose by some chance. If not, there should be something more hidden under the scenes. Actually, this is the case, and taking this `something more' to the forefront involves calling for the octonions. We will give later a constructive approach, and for the time being, we restrict to the purely heuristic approach which was the one adopted at first by Freudenthal.  

\newpage

%_________________________________________________________
\subsection{The `classical' Magic Square}

In the previous section we remarked that there exists only {\itshape four normed division algebras}: the three algebras already considered in the previous constructions $\Re, \Ce, \He$, and just another  member of the family, the octonions $\Oe$. The connection of exceptional simple Lie algebras with the octonions started from the identification, made by Cartan apparently `out of the blue', of the compact form $\g_2^{-14}$  with the Lie algebra $\der\Oe$ of derivations of octonions (which is the Lie algebra of the group of automorphisms of $\Oe$), $\g_2^{-14}\equiv\der\Oe$. 

Now one might think about extending the previous $n$-th floor of the `classical $3\times 3$ Magic Square' by including a fourth column, headed by octonions $\Oe$. To start filling this, one possibility is to build first the spaces whose isometry groups or corresponding Lie algebras should appear in the tables. In the first row, one  should try to construct spheres or better projective spaces over octonions. But unlike the quaternions, where $\He{}P^{n}$ makes full sense and could be constructed for any $n$, there are 
several deep reasons which forbid the existence of projective spaces $\Oe{}P^{n}$ over the octonions for $n>2$. 
However $n=1$ and $n=2$ are exceptions to this impossibility, and the reason is also clear: alternativity meant that product of four octonions involving only three different factors has still some remnant of `associativity'. When trying to establish a would-be $\Oe{}P^2$, instead of taking any possible triplet of octonions as would-be homogeneous coordinates, one may enforce a restriction to take one of the homogeneous coordinates equals to the octonion $\bf 1$, called reduced coordinates. Now in order to make the usual description with homogeneous coordinates to work independently of the chosen charts for these reduced homogeneous coordinates, it turns out that alternativity (and not full associativity) is the only required condition~\cite{Aslaksen}. Hence among the would-be $\Oe{}P^{n}$ with $n\geq2$ only $\Oe P^1$ and $\Oe P^2$ makes actual sense. The plane $\Oe P^2$ is non-desarguesian, as it was discovered by   Moufang in 1933. The actual geometrical reason blocking the possibility of existence of any  $\Oe P^n, n\geq 3$ can be traced to the fact that in such a case, the Desargues property would not be an axiom (as it is for a 2-D projective geometry) but a theorem and  this is in direct conflict with non-associativity of the coordinatising field because a result by Hilbert states that Desargues property can only hold if the coordinate field is associative \cite{Artin}. 

By the 1950s the identifications of the `octonionic special unitary' $3\times 3$ algebras (with two possible signatures) $\Oe \su_3^{0,1}$ to some real forms of the  Lie algebras in the Cartan series $F_4$ was already established \cite{ChevalleySchafer}. These algebras are the isometries of the two octonionic projective planes, the elliptic and hyperbolic Moufang planes. As far as its generators, these include not only the traceless antihermitian octonionic $3\times 3$ matrices, but also the derivations of octonions, with dimension 14.  Indeed the traceless antihermitian octonionic $3\times 3$ matrices span a linear space whose dimension is $2\cdot 7 + 3 \cdot 8 = 38$, yet do not close by themselves a Lie algebra under ordinary matrix commutator, due to the non-commutativity and non-associativity of octonions; if we take into account also the Lie algebra of automorphisms of octonions, with dimension 14 and take a direct sum of these two linear subspaces it is possible to define in this direct sum a structure of Lie algebra, whose dimension is $38+14=52$, which equals the dimension of the Lie algebras in the Cartan series $F_4$. A detailed analysis, counting the number of compact and non-compact generators leads to the precise identification $\Oe \su_3^{0,1} \equiv \f_4^{(-52, -20)}$. Therefore, out of the three possible real forms of the Lie algebras in the Cartan series $F_4$, two of them can be identified with octonionic unitary algebras. 

At this stage, we can arrange in a row the four `unitary' Lie algebras $\Re\su_3^{0,1}$,
$\Ce\su_3^{0,1}$, $\He\su_3^{0,1}$, $\Oe\su_3^{0,1}$ over $\Re, \Ce, \He, \Oe$, which  are the isometries of 2-D projective (spherical) geometry and its $\Ce, \He, \Oe$ relatives. 

\bigskip

{\small
{
\centerline{\vbox{\offinterlineskip
\def\tr{\omit&height 1pt&&&&&&&&}%%% Local \defs, working only inside the group
\def\mr{\omit&height 6pt&&&&&&&&}
\def\myquad{\qquad}
\def\myqquad{\qquad}
\halign{
\strut$#$\quad&\vvrule#\ &\hf$#$\hf&\vrulem#\ &\hf$#$\hf&\vruled#\ &\hf$#$\hf&\vrulem#\ &\hf$#$\hf&\vrulem#\vrulem\cr
           & & \myquad\Re\myquad  & & \myquad\Ce\myquad  & & \myquad\He\myquad   & &  \myquad\Oe\myquad &  \cr
\noalign{\hrule\vskip1pt\hrule}
\mr\cr
    & & \dichar{3}{B_1}{-3,1} & & \dichar{8}{A_2}{-8,0}  &&  \dichar{21}{C_3}{-21,-5}   &&   \dichar{52}{F_4}{-52, -20}  &  \cr\tr\cr
\su    & & \Re\su_3^{0,1} \equiv {\so_3^{\esig 0,1 \esig}}  & & \Re\su_3^{0,1} \equiv{\su_3^{\esig 0,1 \esig}}     &&  \He su_3^{0,1} \equiv{\sq_3^{\esig 0,1 \esig}}   &&  \Oe su_3^{0,1} \equiv {\f_4^{\esig(-52,-20)\esig }}  &    \cr\mr\cr
\noalign{\hrulem}
%\mr\cr
%    & & \dichar{8}{A_2}{2} & & \dichar{16}{2 A_2}{0}  &&  \dichar{35}{A_5}{-7}   &&   \dichar{78}{E_6}{-26}  &   \cr\tr\cr
%\sl   & & \R \sl_3\equiv\sl_3\R & & \C \sl_3\equiv\sl_3\Ce   && \He \sl_3\equiv\su_{6}^* &&  \Oe \sl_3\equiv\e_6^{(-26)} &  \cr\mr\cr
%\noalign{\hruled}
%\mr\cr
%    & & \dichar{21}{C_3}{3} & & \dichar{35}{A_5}{1}  &&  \dichar{66}{D_6}{-6}   &&   \dichar{133}{E_7}{-25}  &   \cr\tr\cr
%\sh   & & \R \sh_6\equiv\sp_{6}\R  & & \C \sh_6\equiv\su_{6}^3  &&  \He \sh_6\equiv\so_{12}^* &&   \Oe \sh_6\equiv\e_7^{(-25)}  &   \cr\mr\cr
%\noalign{\hruled}
%\mr\cr
%    & & \dichar{52}{F_4}{4} & & \dichar{78}{E_6}{2}  &&  \dichar{133}{E_7}{-5}   &&   \dichar{248}{E_8}{-24}  &   \cr\tr\cr
%\msh   & & \f_4^{(4)}  & & \e_6^{(2)}  &&  \e_7^{(-5)} &&   \e_8^{(-24)} &   \cr\mr\cr
\noalign{\hruled}
}}
}
}}

\bigskip
\enlargethispage{\baselineskip}

Now we are being driven by a hope:  that the remaining two families of algebras which we placed in the second and third rows, to wit, $\sl_n$ and $\sh_{2n}$ which had a meaning for geometries over reals, complex and quaternions for all $n$, have also octonionic relatives when $n=3$. Either by careful characterization of the corresponding elements of the Lie algebras as `matrices plus derivations' or by direct geometrical reasoning, one can fill in the two entries for $\Oe\sl_3$ and $\Oe\sh_6$, and perhaps not unexpectedly, these are some real forms of the exceptional Lie algebras in the Cartan series $E_6$ and $E_7$ \cite{Baez:Octonions}. Summing up in a Table all these results, we get: 

\bigskip

{\small
{

\centerline{\vbox{\offinterlineskip
\def\tr{\omit&height 1pt&&&&&&&&}%%% Local \defs, working only inside the group
\def\mr{\omit&height 6pt&&&&&&&&}
\def\myquad{\qquad}
\def\myqquad{\qquad}
\halign{
\strut$#$\quad&\vvrule#\ &\hf$#$\hf&\vrulem#\ &\hf$#$\hf&\vruled#\ &\hf$#$\hf&\vrulem#\ &\hf$#$\hf&\vrulem#\vrulem\cr
           & & \myquad\Re\myquad  & & \myquad\Ce\myquad  & & \myquad\He\myquad   & &  \myquad\Oe\myquad &  \cr
\noalign{\hrule\vskip1pt\hrule}
\mr\cr
    & & \dichar{3}{B_1}{-3,1} & & \dichar{8}{A_2}{-8,0}  &&  \dichar{21}{C_3}{-21,-5}   &&   \dichar{52}{F_4}{-52, -20}  &  \cr\tr\cr
\su    & & \Re\su_3^{0,1} \equiv {\so_3^{\esig 0,1 \esig}}  & & \Re\su_3^{0,1} \equiv{\su_3^{\esig 0,1 \esig}}     &&  \He su_3^{0,1} \equiv{\sq_3^{\esig 0,1 \esig}}   &&  \Oe su_3^{0,1} \equiv {\f_4^{\esig(-52,-20)\esig }}  &    \cr\mr\cr
\noalign{\hrulem}
\mr\cr
    & & \dichar{8}{A_2}{2} & & \dichar{16}{2 A_2}{0}  &&  \dichar{35}{A_5}{-7}   &&   \dichar{78}{E_6}{-26}  &   \cr\tr\cr
\sl   & & \Re \sl_3\equiv\sl_3\Re & & \Ce \sl_3\equiv\sl_3\Ce   && \He \sl_3\equiv\su_{6}^* &&  \Oe \sl_3\equiv\e_6^{(-26)} &  \cr\mr\cr
\noalign{\hruled}
\mr\cr
    & & \dichar{21}{C_3}{3} & & \dichar{35}{A_5}{1}  &&  \dichar{66}{D_6}{-6}   &&   \dichar{133}{E_7}{-25}  &   \cr\tr\cr
\sh   & & \Re \sh_6\equiv\sp_{6}\Re  & & \Ce \sh_6\equiv\su_{6}^3  &&  \He \sh_6\equiv\so_{12}^* &&   \Oe \sh_6\equiv\e_7^{(-25)}  &   \cr\mr\cr
\noalign{\hruled}
%\mr\cr
%    & & \dichar{52}{F_4}{4} & & \dichar{78}{E_6}{2}  &&  \dichar{133}{E_7}{-5}   &&   \dichar{248}{E_8}{-24}  &   \cr\tr\cr
%\msh   & & \f_4^{(4)}  & & \e_6^{(2)}  &&  \e_7^{(-5)} &&   \e_8^{(-24)} &   \cr\mr\cr
\noalign{\hruled}
}}
}

}}

\bigskip

At this point, it is worth to remark that in the second and third rows of the octonionic column, we have just got  only some  of the possible real forms on the complex Lie algebras in $E_6$ and $E_7$; we will come back to this fact later. 

Now one can try the last heuristic step: as the $\Re, \Ce, \He$ initial square was symmetric as far as the Cartan series were concerned, one may try to complete this scheme to a full $4\times 4$ square, which prompts for a complete new fourth row. This called for the invention of   completely new geometries, called metasymplectic by Freudenthal, which could be defined over the reals, complex, quaternions and octonions. The `metasymplectic' geometries are described in some detail in \cite{Freudenthal} (though the description is not easy to decipher at all). 

\bigskip

{\small{
\centerline{\vbox{\offinterlineskip
\def\tr{\omit&height 1pt&&&&&&&&}%%% Local \defs, working only inside the group
\def\mr{\omit&height 6pt&&&&&&&&}
\def\myquad{\qquad}
\def\myqquad{\qquad}
\halign{
\strut$#$\quad&\vvrule#\ &\hf$#$\hf&\vrulem#\ &\hf$#$\hf&\vruled#\ &\hf$#$\hf&\vrulem#\ &\hf$#$\hf&\vrulem#\vrulem\cr
           & & \myquad\Re\myquad  & & \myquad\Ce\myquad  & & \myquad\He\myquad   & &  \myquad\Oe\myquad &  \cr
\noalign{\hrule\vskip1pt\hrule}
\mr\cr
    & & \dichar{3}{B_1}{-3,1} & & \dichar{8}{A_2}{-8,0}  &&  \dichar{21}{C_3}{-21,-5}   &&   \dichar{52}{F_4}{-52, -20}  &  \cr\tr\cr
\su    & & \Re\su_3^{0,1} \equiv {\so_3^{\esig 0,1 \esig}}  & & \Re\su_3^{0,1} \equiv{\su_3^{\esig 0,1 \esig}}     &&  \He su_3^{0,1} \equiv{\sq_3^{\esig 0,1 \esig}}   &&  \Oe su_3^{0,1} \equiv {\f_4^{\esig(-52,-20)\esig }}  &    \cr\mr\cr
\noalign{\hrulem}
\mr\cr
    & & \dichar{8}{A_2}{2} & & \dichar{16}{2 A_2}{0}  &&  \dichar{35}{A_5}{-7}   &&   \dichar{78}{E_6}{-26}  &   \cr\tr\cr
\sl   & & \Re \sl_3\equiv\sl_3\Re & & \Ce \sl_3\equiv\sl_3\Ce   && \He \sl_3\equiv\su_{6}^* &&  \Oe \sl_3\equiv\e_6^{(-26)} &  \cr\mr\cr
\noalign{\hruled}
\mr\cr
    & & \dichar{21}{C_3}{3} & & \dichar{35}{A_5}{1}  &&  \dichar{66}{D_6}{-6}   &&   \dichar{133}{E_7}{-25}  &   \cr\tr\cr
\sh   & & \Re \sh_6\equiv\sp_{6}\Re  & & \Ce \sh_6\equiv\su_{6}^3  &&  \He \sh_6\equiv\so_{12}^* &&   \Oe \sh_6\equiv\e_7^{(-25)}  &   \cr\mr\cr
\noalign{\hruled}
\mr\cr
    & & \dichar{52}{F_4}{4} & & \dichar{78}{E_6}{2}  &&  \dichar{133}{E_7}{-5}   &&   \dichar{248}{E_8}{-24}  &   \cr\tr\cr
\msh   & & \f_4^{(4)}  & & \e_6^{(2)}  &&  \e_7^{(-5)} &&   \e_8^{(-24)} &   \cr\mr\cr
\noalign{\hruled}
}}
}

}}

\bigskip

As one could perhaps expect, the Lie algebras of these real, complex, quaternionic and octonionic metasymplectic geometries are some real forms of the four exceptional series $F_4, E_6, E_7$ and $E_8$.  By playing  numerological relations similar to the ones existing in the previous rows, the characters in the last row can be guessed \cite{Freudenthal}, and through several such tricks, one might expect that the `metasymplectic octonionic' Lie algebra is the real form of the Lie algebra $E_8$ with character $-24$ and dimension $248$. 

This way we find the complete original {\itshape classical Freudenthal Magic Square}. 

Notice that the square is (magically) symmetric as far as Cartan series are concerned, and `magically' because there is nothing in the construction which could suggest such symmetry: columns are labelled by $\Re, \Ce, \He, \Oe$ while rows are labelled by $\su, \sl, \sh, \msh$ which apparently are completely disparate concepts. 

Then, some (but not all real forms of) exceptional algebras in the $F_4, E_6, E_7$ Cartan series are related to the Lie algebras $\su, \sl, \sh$ over octonions, and some (but not all real forms of) exceptional algebras in the $F_4, E_6, E_7, E_8$ Cartan series are related to the Lie algebras $\msh$ of a new geometry  over $\Re, \Ce, \He, \Oe$. 

This completes the original heuristic description of the Magic Square. In the next section we turn over to a more systematic approach, initially developed by Rosenfel'd and which sheds more light on several properties whose appearance in the previous approach comes as by magic. We will also  show that the actual structure behind this Magic Square is a bit  richer than the one just described. 

% >>>>>>>>>>>>>>>>>>>>>>>>>>>>>>>>>>>>>>>>>>>>>>>>>>>>>>>>>>>>>>>>>>

\section{The $\K_1\otimes \K_2$ constructions for the Magic Square} 

Now our aim is to extend the construction of the Magic Square described in the previous section in such a way that provides a `Grand Magic Square' and which accounts for its symmetry as a direct consequence of the construction. Essentially there are two ways to fulfil this aim. We give a rather complete description of one of them and we made a brief reference to the other, which we hope to deal with elsewhere \cite{In prep MS}. 

The first originates from the work by Rosenfel'd in the late 1950s (see full details in \cite{Rosenfeld:GLG}), which was continued by Vinberg (see \cite{Vinberg}). Both authors explored another approach to the Magic Square by looking for the isometry algebras of the 2-D `projective planes' $(\K_1 \otimes \K_2)P^2$ over the tensor product of two composition {\itshape alternative} algebras. One can restrict to take for $\K_1, \K_2$ only the division algebras  $\K_1, \K_2=\Re, \Ce,\He, \Oe$, but the complete approach dictates to  also allow the split versions $\CeS, \HeS, \OeS$.

When compared with the original approach by Freudenthal (or to the more complicated construction by Tits, using Jordan algebras), the Rosenfel'd-Vinberg procedure has a clear advantage, as the symmetry of the Magic Square is somehow built-in in the construction and comes as no surprise at all.
But this is at a price: a purely geometric approach would ask for a previous knowledge of the projective spaces $(\K_1 \otimes \K_2)P^2$ over $(\K_1 \otimes \K_2)$ but details on these `would-be' projective planes over these tensor products are very tricky (and frustratingly not yet clear): nobody knows how to make out sense {\itshape directly} of say $(\OeS \otimes \He)P^2$ or  $(\Oe \otimes \Oe)P^2$ (in a sense similar to the one working for $\Oe P^2$, whose construction is completely well defined). The problem here lies in the fact that tensor products as $\Oe\otimes\He,\ \Oe\otimes\Oe$ are not even alternative. We recall that a `tour de force' machinery was required to deal with projective spaces whose coordinates were possibly non-commutative (as quaternions) or even non-fully associative (as octonions) numbers. In order to properly work, this machinery required the coordinates to belong to an {\itshape alternative algebra}, which is not the case for $(\OeS \otimes \He)P^2$ or  $(\Oe \otimes \Oe)P^2$. Hence, a {\itshape direct} geometrical construction of `would-be' projective planes like $(\OeS \otimes \He)P^2$ or  $(\Oe \otimes \Oe)P^2$ is not available at all. 

Still, there seems to be something definitely right about this $(\K_1 \otimes \K_2)P^2$ sort of ideas, as they lead to amazing heuristic predictions of dimensions, characters, and a lot of arithmetical relations, etc. Then one might try to avoid recourse to the `would-be' projective planes over tensor products as the starting geometric objects and to deal directly with a {\itshape direct} well defined purely algebraic construction of the corresponding Lie algebras. The work by Vinberg follows this line. 

More recently, Barton and Sudbery \cite{BartonSudbery} gave a completely new perspective to understanding the Magic Square, based on the idea of {\itshape trialities} for the composition algebras $\Re, \Ce, \CeS, \He, \HeS, \Oe, \OeS$ and their tensor products. This construction is in some aspects to be preferred over the original by Rosenfel'd and Vinberg, because it puts directly the main emphasis in a property which is likely to be of main importance, the idea of triality. This has tantalizing glimpses of similarities with particle physics; in the Rosenfel'd approach triality stays beyond the scenes. 

In the triality approach the symmetry of the Magic Square is also adequately dealt with, yet the link to  the projective geometries is not clear (which, in view of the difficulties of the approach based on projective planes over tensor products might be perhaps taken as a bonus). 

Now let us to introduce the basic material we will need for the Rosenfel'd-Vinberg construction. 
We make this in several steps. 

% >>>>>>>>>>>>>>>>>>>>>>>>>>>>>>>>>>>>>>>>>>>>>>>>>>>>>>>>>>>>>>>>>>

\section{The construction of the Lie algebra $\K\su_n^l$}

Here $\Ke$ is one of the alternative composition algebras, which is obtained through the CD process, starting from $\Re$ and doubling up to three stages:\quad 
$
\Re \quad \begin{array}{l} \Ce \\ \CeS \end{array} \quad 
\begin{array}{l} \He \\ \HeS \end{array} \quad 
\begin{array}{l} \Oe \\ \OeS \end{array}
$. 
Our first objective is to try to construct the {\itshape special unitary} Lie algebras over $\K$, denoted as $\K\su_n^l$ (or in the $(p,q)$ notation, $\K\su_n^l\equiv\K\su(n-l, l)$). 

Let us first recall the situation in the two cases where $\Ke$ is commutative. It should be clear that the specialization of the general candidate $\K\su_n^l$ for $\Ke=\Re$ and $\Ce$ should reduce directly to $\so_n^l$ and $\su_n^l$. Taking this as the background, we will elaborate the new traits which would be required by the cases $\Ke=\He$ and $\Ke=\Oe$.

Consider first the Lie algebra of the group of isometries of a Hermitian scalar product (with inertia index $l$) over $\Re$. Its metric can be written as a diagonal matrix with $n-l$ positive and $l$ negative terms. To fit the usual CK description, one can as well use the basic CK labelling as introduced in section 2, where the metric can be taken in its diagonal form, 
\begin{equation}
I_n^l \equiv \diag (+,\k_{12},\k_{13},\dots,\k_{1{n-1}}), \qquad 
\k_{1j}:= \k_{2}\k_{3}\cdots\k_{j}, \quad j=2, \dots, n-1,
\label{Metricnl}
\end{equation}
where we recall that the constants ${\k_1, \k_2, \dots, \k_{n-1}}$ are a set of real parameters (the CK labels), which should be non-zero (see more details in \cite{AHPS:CohomSO, HPS:CohomSU, HS:CohomSP,IJTP,SH:GeometOrtogContractions}). The metric does not involve these constants separately, but only their successive products $\k_{1j}$. In the case where all $\k_i$ are different from zero, all the $\k_{1j}$ are also different from zero. If all $\k_i$ are positive, then all the $\k_{1j}$ are also positive, but if there are some negative $\k_i$, then some of the $\k_{1j}$ will be negative. Here $l$ will always denote the number of negative elements $\k_{1j}$ in the diagonal metric matrix; $l$ is precisely the Sylvester `negative inertia index' of the metric. 

The Lie algebra of the group of isometries of a Hermitian scalar product (with inertia index $l$) over $\Re$ (resp. $\Ce$) is the vector space spanned by {\itshape traceless `pseudo-antihermitian' matrices with entries in $\Re$ (resp. $\Ce$)}. We will use the notation 
$\Re\sa_n^l, \Ce\sa_n^l$ for the set of $n\times n$ matrices with entries in $\Re, \Ce$ and which are {\itshape traceless} and  {\itshape pseudo-antihermitian} (pseudo relative to a metric of inertia index $l$); thus for $X\in \Ke\sa_n^l$, we should have the condition:
\begin{equation}
X^T I_n^l  = - I_n^l   X. 
\nonumber
\end{equation}
Notice that for $\Ke=\Re$, `real pseudo-antihermitian' means pseudo-antisymmetric matrices, with ordinary antisymmetry when  $l=0$. For $\Ke=\Ce, \CeS$ we have `complex pseudo-antihermitian'; we must emphasize that our use of `antihermitian' here is a bit imprecise as compared with the normal usage, where it makes reference exclusively to $\Ce$; here it makes implicit reference to the chosen $\Ke$. 
 
The Lie bracket among such generators is given by ordinary matrix commutator. At this point, it is worth to recall that if $A, B$ are two pseudo-antihermitian matrices with entries in either $\Re, \Ce, \CeS$, the matrix commutator $[A, B] = A B - B A$ will be automatically pseudo-antihermitian; this follows by a direct short checking. 
This commutator will be also traceless; this comes from the commutative nature of matrix entries. Hence the ordinary matrix commutator defines a map from $\Ke\sa_n^l \otimes \Ke\sa_n^l$ into $\Ke\sa_n^l$ which automatically  satisfies the Jacobi identity because $\Re$ or $\Ce$ are associative. All this is so familiar that one may easily forget that commutativity or associativity of entries in the matrix is essential to the traceless character of a commutator or the fulfilling of Jacobi identity; while this can be taken for granted for $\Ke=\Re, \Ce$, we now can foresee that this will not work in the same form for $\Ke=\He, \Oe$. 

We must sum up by saying that for $\Ke=\Re, \Ce$, one can construct a Lie algebra denoted $\K\su_n^l$;  at the vector space level $\K\su_n^l$ coincides with the set
$\Ke\sa_n^l$ of $n\times n$ matrices, with entries in $\K$ which are {\itshape traceless} and  {\itshape pseudo-antihermitian} (relative to a metric of inertia index $l$): 
$$
\K\su_n^l =\K\sa_n^l, 
$$
and at the Lie algebra level, the Lie bracket is the matrix commutator: 
$$
[A, B] = A B - B A, \qquad A, B\in \K\sa_n^l.
$$

When applied for $\Ke=\Re, \Ce$, this construction gives the Lie algebras $\so_n^l$ (real entries)  and $\su_n^l$. The replacement of $\Ce$ by its split form $\CeS$ works without any essential change, but there is an important difference in the results: it also leads to a Lie algebra, but here all $\CeS\su_n^l$ for all possible values of $l$ turn out to be isomorphic, something that did not occur for the $\Ce$ case, thus 
the signature index does not play any role in this case. 

Now one may identify which Lie algebra is $\CeS\su_n^l$. Some glimpses are already at hand: it has the same dimension as $\Ce\su_n^l$ and might differ from it only in some sign changes in Lie commutators which come from the replacement from $\Ce$ to $\CeS$; this suggests that $\CeS\su_n^l$ should be another real form of the same complex Cartan family as $\Ce\su_n^l$, and this is actually the case: we have $\CeS\su_n^l\equiv\Re\sl_n$ independently of $l$. A pertinent remark here is that  the view we get on the special linear algebras here is {\itshape rather different from the usual one}; here these appear not as `linear algebras' but instead as `split-unitary' algebras. This new term, `split-unitary', is coined by analogy to `pseudo-unitary' and means `unitary in the sense of split complex numbers'. One may justify this by recalling that similarly we speak of `unitary' for $\su_n$ and `pseudounitary' for $\su_n^l$ with $l=1, 2, \dots$, which is `unitary for a  different Hermitian product with another different signature'. 

We sum up the results up to now: 
$$
\begin{array}{lll}
\Re\su_n^l  \equiv  \so_n^l\equiv \so(n-l, l) &
\qquad \Ce\su_n^l \equiv   \su_n^l\equiv  \su(n-l,l) \cr
& 
\qquad \CeS\su_n^l\equiv\sl_n\Re \cr
\end{array}
$$

But now, what  about the  two remaining and essentially different cases $\Ke= \He, \Oe$? (and their split versions $\HeS, \OeS$?). Is there a construction procedure leading to quaternionic or octonionic unitary Lie algebras $\He\su_n^l,\ \Oe\su_n^l$? Here different new traits appear at each stage, and neither construction is well known, so it is better to discuss this question in two separate steps.  

%_________________________________________________________
\subsection{Extension to $\Ke=\He,  \HeS$ of the construction for  $\K\su_n^l$}

For $\K=\He, \HeS$, the Jacobi identity associated to the
matrix commutator with quaternionic (or split quaternionic) entries still holds (this is {\itshape automatic} whenever $\K$ is
associative). But even if $A, B$ are traceless, then $\tr(A B - B A)$ may be different from zero (however, it is easy to see that $\tr(A B - B A)$ must  be always pure imaginary, so the condition $\text{Re}(\tr(A B - B A))=0$ still holds). Hence the matrix commutator fails to get a Lie algebra structure on the vector space $\K\sa_n^l$ alone, as this matrix commutator produces matrices outside $\K\sa_n^l$. 

To get a Lie algebra starting from this construction, there are two ways out of this failing. One is to discard the `traceless' condition and to enlarge  the $\K\sa_n^l$ vector space accepting into the game all (pseudo-)antihermitian matrices (with vector space denoted $\K\a_n^l$), for which the trace is obviously pure imaginary. Then the matrix commutator of two matrices in $\K\a_n^l$ gives a result which still belongs to $\K\a_n^l$ and as Jacobi identity is ensured, this will close a Lie algebra.
\begin{equation}
\K\su_n^l = \K\a_n^l, \qquad [A, B]= A B - B A. 
\label{RelConmDerHFirst}
\end{equation}

If this construction was done for the complex case, this would lead to the `unitary' (and {\itshape not} the special unitary) Lie algebra $\Ce\u_n^l$, and for this reason the previous result is sometimes stated by saying that there is no a `special unitary' but only a `unitary' quaternionic family of Lie algebras. This choice is not endorsed here, for reasons to be explained in a moment. 

This way out seems to fit well with known facts in the complex case. It turns out that this procedure can be extended for octonions, but only when $n=2$ (where this produces several octonionic interpretations of some orthogonal Lie algebras). Yet there exists another approach for quaternions which allows an extension to octonions in the case $n=3$, and this extension accommodates all exceptional algebras. So let us now first describe it. The starting point is that quaternions, unlike complex numbers, have non-trivial automorphisms, with a Lie algebra of derivations of $\He$ isomorphic to $\so_3$ (derivations for split quaternions are $\so_3^1$). This will mean that further to transformations of space described by matrices acting on coordinates, which will be isometries if suitable conditions are set on these matrices, there will be some additional transformations which come from transforming coordinates themselves by an automorphism. It is clear beforehand that in the real and complex cases there are no any analogue to this situation. Hence the Lie algebra of the group of automorphisms of quaternions must be included in some essential way in the would-be Lie algebra $\He\su_n^l$ (we insist again: if this need did not arise in the $\Re$ or $\Ce$ cases, that was  because for them there are no non-trivial derivations).   

Hence we must enlarge the underlying vector space that we took as our candidate for $\He\su_n^l$ and a new  Lie bracket should be defined in the enlarged space.  
At the vector space level, for $\He\su_n^l$ we take a  direct sum (denoted with the symbol $\ds$ so as to avoid confusion with a direct sum at the Lie algebra level) of the set
$\He\sa_n^l$ of $n\times n$ matrices, with entries in $\He$ which are {\itshape traceless} and {\itshape pseudo-antihermitian} (relative to
a metric of inertia index $l$) {\itshape and} the Lie algebra of derivations of $\K$:
$$
\He\su_n^l = \der\He \ds \He\sa_n^l 
$$
and, at the Lie algebra level, we define the Lie bracket between two elements
of $\He\su_n^l$ in the following form: for $\cal D,
\cal D'\in\der\He$; $A, B\in \He\sa_n^l$, their Lie brackets are defined as 
\begin{equation}
\begin{array}{l}
[\cal D, \cal D']=\cal D \cal D'-\cal D' \cal D,\cr  
[{\cal D}, A]={\cal D} A,\cr 
[A, B]= A B - B A -\frac1n \tr(A B -
B A) \matrixuno + \frac1n {\cal D}(A, B). 
\end{array}
\label{RelConmDerH}
\end{equation}
This candidate  is actually a Lie bracket as it satisfies the Jacobi identity for $\Ke=\He, \HeS$;  here ${\cal D}(A, B)$ is a derivation built from the elements of $A=(a_{ij}), B=(b_{ij})$
\begin{equation}
{\cal D}(A, B):=\sum_{i,j} {\cal D}_{a_{ij},b_{ji}}
\label{DefDerAB}
\end{equation}
where each ${\cal D}_{a_{ij},b_{ji}}$ is the derivation of $\Ke$ given by \myref{DerOctFromCA};  
compare ${\cal D}(A, B)$ with the trace of the product $\tr(AB)=\sum_{i,j} {a_{ij}b_{ji}}$.

For $\K=\Re, \Ce$, the previous identifications $\Re\su_n^l \equiv \so_n^l$ and $ \Ce\su_n^l\equiv\su_n^l$ still fit within the new constructions because in these cases $\der\Re=0, \  \der\Ce=0$ and $\tr(A B - B A) =0$. A relevant comment here is that when applied ---retrospectively--- to the complex case, this construction still provides $\su_n^l$ (and not $\u_n^l$), so this should be looked at as the `special unitary construction' for quaternions. 
In the new cases, $\K=\He, \HeS$, there are non-trivial derivations, whose spaces are $\der\He\equiv \sq_1\equiv \so_3$ for the true quaternions  and $\der\HeS\equiv \sq'_1\equiv \so_3^1$ for the split ones, and their presence is absolutely required in order to have a true Lie algebra for $\He\su_n^l$ or $\HeS\su_n^l$. 

The perceptive reader will probably have noticed that we have given two different constructions for 
the `special unitary algebra over quaternions' \myref{RelConmDerHFirst} and \myref{RelConmDerH}  but we have not even bothered to use different names for them. This is intencional, because what happens is that in spite of the superficially more complicated Lie brackets in \myref{RelConmDerH} for a quaternionic algebra $\He, \HeS$, it turns out that the two Lie algebras arising from both constructions are actually isomorphic. 
Indeed if we are interested only in the quaternion case, there is a natural identification among pure imaginary unit quaternions and the derivations of $\He$ (both close Lie algebras which are isomorphic, and the same happens for $\HeS$). In complete detail, this is
a consequence of relation ${\cal D}_{\vx, \vy}=C_{[\vx, \vy]}$ between derivations and commutator maps valid for $\He, \HeS$ (and, indeed but trivially, for $\Re, \Ce, \CeS$ also). Hence the derivations, acting on an equal basis on each coordinate, can be secretly seen through matrices which are pure imaginary multiples of the identity matrix (antihermitian of course yet not traceless). Within this interpretation the derivations are out of sight and the Lie algebra $\K\su_n^l$ can be described simply as spanned by (pseudo)-antihermitian quaternionic matrices (with a metric of signature index $l$) whose trace has real part equal to $0$ and with simple matrix commutator as Lie bracket. 

In this quaternionic case this double interpretation depends on the possibility of a complete `quaternionic trading' among the two terms ${\cal D}(A, B)$ and $\tr(A B - B A)$, but this is a specifically quaternionic trait, so from a general perspective it is much better to stick only to the second construction \myref{RelConmDerH} which at the end will be the one working ---only for $n=3$--- also for octonions. 

We sum up the results up to now by saying that the three main `signature' series of simple Lie algebras as well as $\sl_n\Re$ and $\sp_n\Re$ appear as `unitary' algebras over $\Re, \Ce, \CeS, \He, \HeS$ as described in the display
$$
\begin{array}{lll}
\Re\su_n^l  \equiv  \so_n^l\equiv \so(n-l, l) &
\quad \Ce\su_n^l \equiv   \su_n^l\equiv  \su(n-l,l) &
\quad \He\su_n^l \equiv   \sq_n^l\equiv \sq(n-l,l) \cr
& 
\quad \CeS\su_n^l\equiv\sl_n\Re &
\quad \HeS\su_n^l\equiv\sp_{2n}\Re \cr
\end{array}
$$

The new construction \myref{RelConmDerH}  turns $\Re\su_n^l,\, \Ce\su_n^l,\, \CeS\su_n^l,\, \He\su_n^l,\, \HeS\su_n^l$ into Lie algebras.  For all these Lie algebras, it is a quite simple exercise to check the identifications as stated, and to compute its dimensions, characters, etc. directly from the realizations in terms of traceless $\K$-antihermitian matrices and derivations of $\K$. 

%_________________________________________________________
\subsection{Extension to  $\K=\Oe, \OeS$ of the construction for  $\K\su_n^l$: the exceptional case $n=3$}

Now we arrive to the most complicated stage: the octonionic case. The Lie algebras of derivations of $\K=\Oe, \OeS$ are 
$
\der\Oe\equiv \g_2^{(-14)},\ 
 \der\OeS\equiv \g_2^{(2)}\  
$.
We also recall the relation ${\cal D}_{\vx, \vy}=C_{[\vx, \vy]}-3 A_{\vx, \vy}$ linking derivations for $\K=\Oe, \OeS$ with the corresponding commutator and associator maps; the coefficient $3$ in the associator term here is fixed, and only appears for octonions (or their  split versions), as associators are identically zero for all the remaining composition algebras. 

Derivations are antisymmetric maps, so the derivations of the octonions or split octonions should belong to the Lie algebras $\so(\Oe)$ or $\so(\OeS)$, which are isomorphic to $\so_8, so_8^4$, with dimension $28$ (here $\so(\Oe)\equiv \so_8$ denotes the algebra of the orthogonal group behind the octonions seen as an 8-D vector space with its natural positive definite inner product; for split octonions, $\so(\OeS)\equiv \so_8^4$). 
As every derivation annihilates the octonion $\mb 1$, then it is clear that $\der\Oe, \, \der\OeS$ sit inside the subalgebras 
$\so(\ImPart\Oe)\equiv\so_7,\  \so(\ImPart\OeS)\equiv\so_7^3$ which have dimension $21$, and this means that the derivations of octonions, with dimension 14, are only a subset of $\so(\ImPart\Oe),\  \so(\ImPart\OeS)$ (in clear distinction to the quaternionic case, where actually we had an equality $\der\He = \so(\ImPart\He)$). There are 21 octonionic associators $\CalA_{\ve^\alpha, \ve^\beta}$ with $\alpha<\beta$, and all of these can be  checked to be indeed linearly independent, so it turns out that octonionic associators do close a Lie algebra isomorphic to $\so(\ImPart\Oe),\  \so(\ImPart\OeS)$, and by taking into account the relation \myref{DerOctFromCA}, then it follows that these Lie algebras are also spanned by the octonionic derivations together with the octonionic commutator maps.

Now we ask whether the $\K\su_n^l$ construction can be made to work for $\Ke=\Oe, \OeS$. For matrices with octonionic entries, again the matrix commutator of two traceless matrices is not traceless (though the trace is still pure imaginary). This failing was already found at the quaternionic level, where we gave two procedures to way out. Mimicking the first one, one can try to relax the traceless condition for (pseudo)-antihermitian octonionic matrices in a way similar to the one made in \myref{RelConmDerHFirst}. For octonions however, things cannot be made to work in such a simple way due to the lack of full associativity of octonions: it suffices to see that the set of (scalar multiples of the unit matrix whose  coefficients are) the seven pure imaginary octonionic units do not span a Lie algebra, but a slightly weaker structure, called a Moufang loop. Hence this naive attempt does not work in the same way it did for quaternions, though some modifications which also involve adding the derivations can be made to work in the $n=2$ case with $\Ke=\Oe$. 

Now we might turn to the other procedure used for $\He$, and try to take in the octonionic $\Oe, \OeS$ case a `would-be' Lie algebra   
$\Oe\su_n^l,\, \OeS\su_n^l $ as a direct sum of the vector spaces of derivations of $\Oe, \OeS$ and of the set of traceless antihermitian matrices with entries in $\Oe$:
$$
\K\su_n^l:=\der\K\ds\K\sa_n^l, \qquad \Ke=\Oe, \OeS, 
$$
with would-be Lie brackets given by \myref{RelConmDerH} and \myref{DefDerAB}. In all the previous cases with $\Ke$ real, complex or quaternionic, this candidate actually satisfied Jacobi identity, and hence defined a Lie algebra. Now what happens is that when $\Ke$ is octonionic, for any general $n$ then \myref{RelConmDerH} {\itshape does not satisfy Jacobi identity}, but there is an {\itshape exceptional case} where the Jacobi identity is actually satisfied: the exceptional case is $n=3$. Where the exceptional behaviour for $n=3$ comes from? It turns out that in order to \myref{RelConmDerH} should satisfy Jacobi identity, one has to have a careful balance between three numbers which appear in the construction: the $n$ in the $\frac1n$ appearing as the coefficient in $ {\cal D}(A, B)-\tr(A B -B A) \matrixuno$, the $n=\tr\matrixuno$ and a fixed $3$ appearing in \myref{DerOctFromCA}, which comes from the description of derivations of octonions (of course, this pure number $3$ is related to triality and only appear whenever $A_{\vx, \vy}\neq 0$, but the important point here is that this last $3$ {\itshape cannot} be changed, as it is an intrinsic property of octonions). 

Hence, the only case of \myref{RelConmDerH} which gives a Lie algebra when $\Ke$ is octonionic is $n=3$ and in this case the Lie brackets are  
\begin{equation}
\begin{array}{l}
[\cal D, \cal D']=\cal D \cal D'-\cal D' \cal D ,\cr  
[{\cal D}, A]={\cal D} A ,\cr 
[A, B]= A B - B A -\frac13 \tr(A B -
B A) \matrixuno + \frac13 {\cal D}(A, B) ,
\end{array}
\label{VinbergCandidate3}
\end{equation}
where ${\cal D}(A, B)$ is given by the same expression as in \myref{DefDerAB}
%
%Therefore, in the octonionic cases where the associators are different from zero, this construction provides a Lie algebra only for $n=3$. This is consistent with a fact we mentioned before, that reasons reasons coming from {\itshape projective geometry} stated it was impossible to obtain a projective geometry with more that two octonionic dimensions (whose Lie algebra of isometries would be a such $\Oe\su_n$ with $n>3$); if this were the case, Desargues would be a theorem, not an axiom, and this is precluded because Desargues only holds if the coordinate field is  associative. 
%

The attentive reader will undoubtedly have noticed that while the projective space $\Oe P^1$ also exists, its would-be algebra does not appear within the Rosenfel'd-Vinberg theorem; we will come back on this question in  a moment.  

Now numerology counts for $\K=\Oe, n=3$ can be done quite easily, and guides the identification of the Lie algebras  thus obtained. As a simplest example, in a traceless $3\times 3$ antihermitian matrix with octonionic entries there are ${3\choose2} = 3$
off-diagonal matrix octonionic entries (each amounting to 8 real numbers) and 2 diagonal pure imaginary octonionic entries (each amounting to 7 real numbers). Adding the 14 derivations of $\Oe$ one gets a grand total of $3 \cdot 8 + 2 \cdot 7 + 14 = 52$ which is precisely the dimension of the exceptional algebra $\f_4$. Similar numerology counts can also be performed  for the characters. 
In this way, the algebras $\Oe\su_3^l$ are identified to two of the real forms of the exceptional Lie algebra $\f_4$ and for $\K=\OeS$ this construction gives also the third real form of the same exceptional complex Lie algebra $\f_4$ as the `split unitary octonionic algebra'. Hence, we can sum up this by saying that all the three real forms of the exceptional Lie algebra $\f_4$ appear as `octonionic special unitary' algebras, completing the previous displays with a last entry:  
$$
\begin{array}{llll}
\Re\su_n^l \!\equiv\! \so_n^l &
\qquad \Ce\su_n^l\!\equiv\!  \su_n^l &
\qquad \He\su_n^l\!\equiv\!  \sq_n^l &
\qquad \Oe\su_3^{0,1} \equiv \f_4^{(-52, -20)}\cr
& 
\qquad \CeS\su_n^l\equiv\sl_n\Re &
\qquad \HeS\su_n^l\equiv\sp_{2n}\Re & 
\qquad \OeS\su_3^{0,1} \equiv \f_4^{(4)} 
\end{array}
$$

Of course, because there are no traceless matrices in $n=1$, this construction already works in a rather trivial form when $n=1$ for any $\K$  suggesting to {\itshape define} 
$$
\begin{array}{ll}
\Re\su_1\equiv \der\Re=0\cr
\Ce\su_1\equiv \der\Ce=0 \qquad&\CeS\su_1\equiv \der\CeS=0  \cr
\He\su_1\equiv \der\He=\sq_1\equiv\so_3\qquad &\HeS\su_1\equiv \der\HeS=\sq'_1\equiv\so_3^1\cr
\Oe\su_1\equiv \der\Oe=\g_2^{(-14)}\qquad &\OeS\su_1\equiv \der\OeS=\g_2^{(2)}\cr
\end{array}
$$
which would make the two real forms of the Cartan series $\g_2$ to appear. 

Up to this point we have described an interpretation of all the real forms in the Lie algebras of the Cartan series $\g_2$ and $\f_4$.  

%_________________________________________________________

\section{The defining construction of the Lie algebra $(\K_1\otimes\K_2)\su_n^l$ and the Grand Magic Square}

The previous construction is only a preparation to our main goal, which is to give a different construction of the Magic Square which would provide some clue to its properties, in particular to its symmetry (in the original Freudenthal approach, the symmetry was only as far as Cartan series were concerned). Rosenfel'd proposed to approach the Magic Square in a way which is much more symmetric than the original Freudenthal construction (Tits construction was also quite complicated, involving Jordan algebras). Rosenfel'd idea was to label rows and columns of the square by two division algebras $\K_1, \K_2$ and heuristically understand each entry as a `special unitary' algebra over $(\K_1\otimes\K_2)$. In 1966 Vinberg gave an explicit and direct construction of these {\itshape special unitary} Lie algebras over the tensor products $\K_1\otimes\K_2$. This  follows closely the construction presented in the previous subsections, whose aim has actually been to pave the way for this final construction. 

At the vector space level, $(\K_1\otimes\K_2)\su_n^l$ is taken as the linear direct sum of {\itshape derivations} of $\K_1\otimes\K_2$ plus the vector space of {\itshape traceless (pseudo)antihermitian} $n\times n$ matrices with entries in $\K_1\otimes\K_2$ (relative to a metric which possibly has signature with inertia index $l=0,1$) 
\begin{equation}
(\K_1\otimes\K_2)\su_n^l = \der(\K_1\otimes\K_2) \ds (\K_1\otimes\K_2)\sa_n^l.
\label{RelCommVinbergKKSpace}
\end{equation}

This requires to know the Lie algebra of derivations of a tensor product, which according to the requirement of Leibniz property should necessarily be defined in terms of the derivations ${\cal D}_{\vx_1, \vy_1}$ and ${\cal D}_{\vx_2,\vy_2}$ on each factor by means of:
$$
{\cal D}_{\vx_1\otimes\vx_2,\ \vy_1\otimes\vy_2}:=
\HermProd{\vx_1}{\vy_1}{\cal D}_{\vx_2,\,\vy_2} +  
\HermProd{\vx_2}{\vy_2}{\cal D}_{\vx_1,\, \vy_1}, 
$$
which means that up to isomorphism
$$
\der(\K_1\otimes\K_2) \equiv \der(\K_1) \oplus \der(\K_2).
$$

The Lie brackets in the `would-be' Lie algebra $(\K_1\otimes\K_2)\su_n^l$ are just the ones in \myref{RelConmDerH}.

For this construction, the final result is similar to the previous one: 
the vector space $(\K_1\otimes\K_2)\su_n^l$  \myref{RelCommVinbergKKSpace} with the  brackets \myref{RelConmDerH} closes a Lie algebra in the following cases: 

\begin{itemize}

\item
For all $n=2, 3, \dots$ and both $\K_1, \K_2$ in the list $\Re,\Ce,\CeS,\He, \HeS$. 

\item
Exceptionally {\itshape only for $n=3$}, when  $\K_1, \K_2$ are in the list $\Re$, $\Ce$, $\CeS$, $\He$, $\HeS$,  $\Oe$, $\OeS$ and at least one of the $\K_1, \K_2$  is octonionic. 
\end{itemize}

The proof consists in checking that Jacobi identities are actually satisfied; this depends on the combined {\itshape alternative and composition} character of the factors $\K_1, \K_2$ and the fact that $\K_1\otimes\K_2$ itself is not alternative does not matter. And again the restriction to $n=3$ when either $\K_1, \K_2$ is octonionic comes from the reason discussed when we presented the $\K\su_n^l$ case in the previous pages. 

Of course, the construction also works (though this case is trivial) for $n=1$, where the set of traceless $1\times 1$ matrices is zero-dimensional, and here only the part coming from the derivations survives. This suggests to {\itshape define} 
$$
(\K_1\otimes\K_2)\su_1 =
\der(\K_1\otimes\K_2). 
$$

Now we can see in which aspects this new construction sheds more light than the original `Freudenthal' one previously described. To start with, this approach  actually suggests not only one but {\itshape three} different Magic Squares, which we could call the `compact' one, the `mixed' one and the `split' one. These are obtained  by successively selecting for $\K_1, \K_2$ the following three choices: 
\begin{equation}
\begin{array}{llll}
&\K_1=\Re, \Ce, \He, \Oe, &\quad \K_2=\Re, \Ce, \He, \Oe, &\quad \hbox{The `compact' Magic Square.\hfill\ }\nonumber\cr
&\K_1=\Re, \CeS, \HeS, \OeS, &\quad \K_2=\Re, \Ce, \He, \Oe, &\quad \hbox{The `mixed' Magic Square.}\nonumber\cr
&\K_1=\Re, \CeS, \HeS, \OeS, &\quad \K_2=\Re, \CeS, \HeS, \OeS, &\quad \hbox{The `split' Magic Square.}\nonumber
\end{array}
\end{equation}

Instead of displaying all the three $4\times 4$ Magic Squares (with inertia indices appearing effectively only when both algebras $\K_1, \K_2$ are division ones), it is perhaps better to arrange all the algebras in a single `Grand Magic Square', whose box-entries are the Lie algebras denoted  $(\K_1\myot\K_2)\su_n^l$ coming  from the Rosenfel'd-Vinberg construction.  We first display this {\em Grand Magic Square} for $n=3$ and then will comment on some aspects. 

\bigskip

{{

\centerline{\vbox{\offinterlineskip
\def\mr{\omit&height 6pt&&&&&&&&&&&}%%% Local \defs, working only inside the group
\def\myquad{\quad\!\!}
\def\myqquad{\qquad}
\halign{
\strut$#$\quad&\vvrule#\ &\hf$#$\hf&\vrulem#\ &\hf$#$\hf&\hf$#$\hf&\vruled#\ &\hf$#$\hf&\hf$#$\hf&\vrulem#\ &\hf$#$\hf&\hf$#$\hf&\vruled $#$\cr
           & & \myquad\Re\myquad  & & \myquad\Ce\myquad  & \myquad\CeS\myquad  & & \myquad\He\myquad  & \myquad\HeS\myquad  &&  \myquad\Oe\myquad  & \myquad\OeS\myquad\      &  \cr
\noalign{\hrule\vskip1pt\hrule}
\mr\cr
\Re    & & {\so_3^{\esig 0,1 \esig}} & & {\su_3^{\esig 0,1 \esig}}  & {\sl_3\Re}   && {\sq_3^{\esig 0,1 \esig}}   & {\sp_{6}\Re} &&{\f_4^{\esig(-52,-20)\esig }}& {\f_4^{(4)}}   &  \cr\mr\cr
\noalign{\hrulem}
\mr\cr
\Ce   & & {\su_3^{\esig 0,1 \esig}} & & {\Dos\su_3^{\esig 0,1\esig}}& {\sl_3\Ce}   && {\su_{6}^{\esig 0,2 \esig}}& {\su_{6}^3}  && {\e_6^{\esig(-78,-14)\esig}}& {\e_6^{(2)}}   &  \cr\mr\cr
\CeS   & & {\sl_3\Re}                 & & {\sl_3\Ce}                   &{\Dos\sl_3\Re}&& {\su_{6}^*}                & {\sl_{6}\Re}  &&{\e_6^{(-26)}} & {\e_6^{(6)}}&  \cr\mr\cr
\noalign{\hruled}
\mr\cr
\He   & & {\sq_3^{\esig 0,1 \esig}} & & {\su_{6}^{\esig 0,2\esig}} & {\su_{6}^*}  &&{\so_{12}^{\esig 0,4 \esig}}& {\so_{12}^*} && {\e_7^{\esig(-133,-5)\esig}}& {\e_7^{(-5)}}    &  \cr\mr\cr
\HeS   & & {\sp_{6}\Re}               & & {\su_{6}^3}                & {\sl_{6}\Re}  && {\so_{12}^*}               &{\so_{12}^{6}}&& {\e_7^{(-25)}}              & {\e_7^{(7)}}   &  \cr\mr\cr
\noalign{\hruled}
\mr\cr
\Oe   & &{\f_4^{\esig(-52,-20)\esig}}& &{\e_6^{\esig(-78,-14)\esig}}&{\e_6^{(-26)}}&&{\e_7^{\esig(-133,-5)\esig}}&{\e_7^{(-25)}}&&{\e_8^{\esig(-248,-24)\esig}}& {\e_8^{(-24)}} &  \cr\mr\cr
\OeS   & & {\f_4^{(4)}}               & & {\e_6^{(2)}}               & {\e_6^{(6)}} &&{\e_7^{(-5)}}               & {\e_7^{(7)}}  && {\e_8^{(-24)}}              & {\e_8^{(8)}}   &  \cr\mr\cr
\noalign{\hruled}
}}
} 

}}

\bigskip

There are three remarkable things here.  

First, once the `ordinary' as well as the split versions of reals, complex, quaternions and octonions are considered for both $\K_1$ and $\K_2$ in the construction, {\itshape all the real forms for the exceptional algebras} in the four series $F_4, E_6, E_7, E_8$ appear here as `special unitary' Lie algebras $(\K_1\,\myot\,\K_2)\su_3^l$ when at least one of the two alternative composition algebras $\K_1$, $\K_2$ is octonionic. For instance, the split form $\fivs$ of the Cartan series $\f_4$ did not appear in the original Freudenthal Magic Square, but if we look to the Grand Magic Square, we find that $\fivs$ appears there as $(\Re\otimes\OeS)\su_3^l$, i.e., as the `split-unitary' octonionic algebra. A cursory look to the Grand Magic Square shows that the same happens for all real forms of the exceptional algebras in the Cartan series $\e_6$, $\e_7$ and $\e_8$: all the three real forms of $\f_4$, all the five of $\e_6$, all the four of $\e_7$ and all the three of $\e_8$ appear. 

Second, this Grand Magic Square is completely symmetric, as far as the real forms of Lie algebras are concerned. This means complete symmetry   around the main diagonal for {\itshape both the Cartan series and for the characters} (and not merely symmetric as far the Cartan series were concerned, as it was in the original Freudenthal Magic Square). The full symmetry of the Magic Square shows up clearly in its Grand version, where we consider altogether the `compact' and the `split' composition alternative algebras, giving rise to three interleaved `partial Magic Squares'. 

Third,  it is clear that the original Freudenthal Magic Square is just the `mixed' subsquare of the Grand Magic Square obtained by taking for $\K_1$ the `split' versions $\K_1=\Re, \CeS, \HeS, \OeS$ of the composition alternative algebras, while for $\K_2$ the division ones $\K_2=\Re, \Ce, \He, \Oe$ are taken.  The fact that this was symmetric only as far as the Cartan series but not for characters should be traced back to the fact that this is a non-symmetric subsquare of the full Grand Magic Square. 

%_________________________________________________________
\subsection{Are there any extension to other   Grand Magic Squares for $n\neq3$?}

This is a quite natural question, and a part of the answer can be easily guessed: for $n>3$ if we let octonions enter as $\K_1$ or $\K_2$ there is no any possibility of closing a Lie algebra $(\K_1\otimes\K_2)\su_n^l$; the geometric reason already mentioned provides a rather final impossibility condition: this is linked with the fact that the Desargues property in $n>3$ would require a coordinatising field which should be associative. 

For $n=1$ we would have $1\times 1$ traceless matrices with entries in $(\K_1\otimes\K_2)$ as a part of the construction \myref{RelConmDerH}, and as the set of such matrices contains only the $0$ matrix, we get a rather trivial result: for $n=1$, $(\K_1\otimes\K_2)\su_1$ should refer to the Lie algebra of derivations of $(\K_1\otimes\K_2)$ alone, which is a direct sum $\der \K_1 \oplus \der\K_2$. This was mentioned in passing when we introduced the Classical Magic tower, where the base floor, with $n=1$ contains some Lie algebras, as $\He\su_1$, which are not trivial (and now we see that for all the entries, the $\K\su_1$ algebra should indeed coincide with the derivations of $\K$, which are trivial for $\K=\Re, \Ce, \CeS$ and non-trivial for $\K=\He,\HeS, \Oe, \OeS$). 

But still there remains the case  $n=2$, which would correspond to the projective space $\Oe P^1$. 
For $n=2$, things are a bit more tricky. By one side, further to the  projective spaces $\Ce P^1$ and $\He P^1$, it is clear that $\Oe P^1$ is a well defined object, both as a subgeometry of $\Oe P^2$  and as a direct construction by using homogeneous coordinates, a construction which works for $\Oe P^1$ by the same reason as it worked for $\Oe P^2$ (as in this case would-be homogeneous coordinates are given by two different octonions, and there are only two sets of reduced homogeneous coordinates for each point, whose compatibility is easy to proof; recall that reduced meant that one of the coordinates is taken as the octonion $\bf 1$). Hence, one should expect a well defined Lie algebra of isometries there. The only thing which is clear beforehand is that when octonions appear, this algebra {\itshape cannot be given} by \myref{RelConmDerH}  with $n=2$, because this is not even a Lie algebra and the Jacobi identity fails for it.

To understand what is happening here, let us go back to the Lie algebra $\su_3$, in its ordinary matrix realization with $3\times 3 $ matrices. This contains naturally three copies of the Lie algebra $\u_2$ (and not $\su_2$), essentially because a $3\times 3 $  matrix which was traceless contains three $2\times 2$ diagonal submatrices which are no longer neccessarily traceless. And of course this is linked to the standard realization of $\u_2$ as a sum of $\su_2$ and a 1-D subalgebra, isomorphic to $\u_1$, which is generated by the scalar multiples of the identity $2\times 2$ matrices with $i$ as the coefficient. The point here is that the `complex unitary algebras' can be obtained from the `special unitary ones' by adding {\itshape a single generator}, whose matrix is identified to the imaginary unit (this closing trivially a 1-D Lie algebra). Needless to say, while $\su_n$ is a simple Lie algebra, $\u_n$ is not. 

Are there a construction for `unitary algebras' over $\K$ or over $\K_1\otimes\K_2$ extending this construction? Let us mention that the literature at this point can easily lead to some confusion. Consider the quaternion case, $\K=\He$.  The `quaternionic special unitary Lie algebra' was generated by traceless antihermitian matrices and the derivations of quaternions. But there is a coincidence: the Lie algebra of derivations for quaternions is isomorphic to the Lie algebra closed by the three  quaternionic imaginary units, which is clearly isomorphic to $\so_3$. This means that some trading is possible by mimicking the role of derivations in the special unitary Lie algebra by matrices which are Hermitian but not traceless, and which are scalar multiples of the identity matrix with a pure imaginary quaternion as coefficient. It turned out that the set of (pseudo)-antihermitian matrices with quaternionic entries (discarding the traceless condition) did actually close a Lie algebra, and this was related to the fact that the pure imaginary quaternions did close by themselves a Lie algebra. 

When we go to the octonionic case, things are again different. In this case the $n=3$ `octonionic special unitary Lie algebra' is generated by $3\times 3$ traceless antihermitian matrices and the derivations of octonions, which altogether do close a Lie algebra, while the seven  octonionic imaginary units alone do not close a Lie algebra. This means that  in the $n=3$ octonionic `special unitary algebra' there is not any similar possibility of trading among derivations and pure imaginary units as it was possible with quaternions. Here somehow the octonions remind the situation in the complex case: for $\Ce$ there are no derivations, so no possibility of trading among derivations and the pure imaginary unit), but for complex numbers the algebras $\Ce\su_n^l$ and $\Ce\u_n^l$ are different objects, which exist for all $n$, with $\Ce\u_n^l$ obtained from $\Ce\su_n^l$ by adding a single generator (realized by the matrix $i\matrixuno$) and only $\Ce\su_n^l$ being simple.

After these comments, we can give the end result: further to the `special unitary octonionic' construction of a Lie algebra $\Oe\su_3^l$ which works only for the precise value $n=3$, one can build a  `unitary octonionic' Lie algebra $\Oe\u_2^l$ through a construction which works {\itshape only for $n=2$}. We cannot go into full details here for lack of space, because some details are tricky, but we mention the basic idea:  using \myref{DerOctFromCA} one can replace derivations in terms of commutator and associator maps, and in the would-be Lie brackets among traceless matrices in \myref{RelConmDerH}, there is a cancellation among the commutators coming from the derivations and the trace part, which in the quaternionic case was complete, but in the octonion one is only partial, as the associators coming from the derivations are still present. This means that even if one accept the antihermitian octonionic matrices, relaxing the traceless requirement, then a good Lie bracket will not be provided by a matrix commutator alone, as the associator maps should also appear. Then a careful study discloses that even starting from $\Oe\a_n^l$, 
(which should lead to the `unitary' algebra, not to the `special unitary' one), one has still to add the associators. But there is a natural identification of octonionic associators with a direct sum of `imaginary octonions' plus derivations. As a result of the existence of these linear  relations, at the end of the day the Lie brackets are a bit more complicated than the ones in  \myref{RelConmDerH}. This idea however can be taken in some heuristic manner to compute dimensions and characters: as far as a linear vector space, $\Oe\u_2^l$ should be a direct sum of the antihermitian traceless $2\times2$ matrices (whose dimension is 15), the antihermitian scalar multiples of the identity matrices (not traceless, with dimension 7) and the derivations of octonions (dimension 14). This gives a grand total of $36$, and it turns out that the Lie algebras $\Oe\su_2^l$ are isomorphic to $\so_9^{0,4}$. This is the first example of a classical Lie algebra which further to its `natural' realization in terms of real numbers has also an octonionic realization. 

We remark that this allows to look to this new construction as a kind of variant of the Vinberg one, working only for $n=2$, where one adds to $\Oe\sa_2^l$ (dimension 15) the full Lie algebra of associators (which is isomorphic to $\so_7$, with dimension 21) getting a Lie algebra with dimension $36$; the fact that some choice of seven linear combinations of the 21 associators can be `identified' to the seven octonionic units justifies to call the Lie algebra thus obtained as `unitary' and not as  `special unitary'.

As far as we know, the first systematic discussion of this construction was done by Sudbery \cite{Sudbery84}, who related the octonionic members of this family to some orthogonal Lie algebras and their associated spinors. Another reason for not providing full details here is that there is another construction of the exceptional Lie algebras, based on the concept of triality, first given by Barton and Sudbery \cite{BartonSudbery} which turns out to provide directly and in a single run the three families $(\K_1\otimes\K_2)\su_3^l$,  $(\K_1\otimes\K_2)\u_2^l$ and $(\K_1\otimes\K_2)\su_1$. 

In the last section, we include full tables with the octonionic identifications of the orthogonal Lie algebras. 

%_________________________________________________________
\subsection{Can all simple real Lie algebras be seen as `antihermitian'?}

The `special unitary' Lie algebras $(\K_1\myot\K_2)\su_3^l$ we have constructed in the special case $n=3$ are intended to correspond  to a `projective' geometry on the 2-D `projective planes' $(\K_1 \otimes \K_2)P^2$, which should be locally identical to the corresponding spherical geometries. But we should not forget that while the construction of these Lie algebras is well defined, the idea of the 2-D `projective planes' $(\K_1 \otimes \K_2)P^2$ has at most some heuristic meaning. There is something rather frustrating when we realize that after more than half a century, we still do not know how to make precise sense out of this idea which is undoubtedly `correct' at some heuristic level. Then the prevailing viewpoint is to take seriously the `vague' connections only as far as they can be supplemented with well defined algebraic constructions for the Lie algebras themselves.  

The intelligence of the the exceptional Lie algebras was after all the main drive in all this search. We can now say that all exceptional simple Lie algebras can actually be seen as `special unitary algebras' $(\K_1\otimes \K_2)\su_3^l$, not forgetting that $(\K_1 \otimes \K_2)P^2$ has only some heuristic meaning when $\K_1$ or $\K_2$ is octonionic. When both $\K_1$ and $\K_2$ are taken among the division algebras $\Re, \Ce, \He, \Oe$, then there is an inertia index $l=0,1$ which serves to distinguish among two of the possible real forms (as exemplified by the case, $\Oe\su_3^l\equiv\f_4^{(-52,-20)}$), and in the tables we display either the possible inertia indices (in superscript position, without parentheses) or the possible characters (in superscript position, inside parentheses). The remaining exceptional algebras in the $\g_2$ family, also appear as the $(\K_1\myot\K_2)\su_1$ family. 

We can sum up these results is a nutshell by saying that {\itshape all exceptional Lie algebras are `special unitary'}; this actually means that all exceptional simple Lie algebras can be understood as the result of the construction described before with suitable choices for $\K_1$ and $\K_2$ for $n=1, 3$, including at least one $\K_1$ or $\K_2$ octonionic,  but also some other classical algebras appear in these cases when neither $\K_1$ or $\K_2$ is octonionic. Indeed, a related construction is possible also for $n=2$ leading to algebras $(\K_1\myot\K_2)\u_2^l$ which are certain and classical algebras even if $\K_1$ or $\K_2$ are octonionic. The `ingredients' in the construction are: 

\begin{itemize}

\item 
A fixed value $n=2, 3$, which sets the order of the matrices in the linear space of matrices $(\K_1\otimes\K_2)\sa_n$. 

\item 
A choice of two alternative composition algebras $\K_1, \K_2$; the choice is restricted to $\Re, \Ce, \CeS, \He, \HeS$ for any $n$ and exceptionally $\Oe, \OeS$ are allowed for either $\K_1, \K_2$ only when $n=2,3$. 

\item 
Finally, a metric matrix for a `Hermitian' (relatively to the $*$-conjugation antiinvolution)  form in the space $(\K_1\otimes\K_2)^n$, which can be taken  diagonal with non-zero real values $(1, \k_{12},\k_{13},\cdots\k_{1n-1})$ which altogether describe a metric with inertia index $l$, which equals the number of negative terms in the diagonal, in the same fashion as described in the $\so_{\k_1 \k_2\cdots \k_{n-1}}(n)$ family of CK algebras (for the simplest cases, see \cite{AHPS:CohomSO,HPS:CohomSU,HS:CohomSP}). 

\item 
Exceptionally, $n=1$ is also allowed, and in that case the antihermitian construction, either special unitary or unitary, gives the Lie algebra of derivations of $\K_1\otimes\K_2$. 
\end{itemize}

%_________________________________________________________

\section{Non-standard antiinvolutions in $\Ce, \CeS, \He, \HeS$}

After having checked that all the {\itshape exceptional} simple real Lie algebras can be understood as $(\K_1\otimes\K_2) \su_3^l$ or  $(\K_1\otimes\K_2) \su_1$ with an octonionic $\K$, one may set a question: are {\itshape all} simple real Lie algebras `special unitary' in  a similar sense? Within the context we have described up to now, the answer is only {\itshape almost}. Some simple real Lie algebras  definitely do not not appear in the previous construction: two instances which do not admit such realization are $\so_n\Ce$ and $\so_{2n}^*$ for $n$ odd ---in the case $n$ even, $2n=4n'$ is a multiple of four and the algebra $\so_{2n}^*\equiv\so_{4n'}^*$ can be understood as `special unitary' over $\He\otimes\He$---. 

Thus a natural question is: can we devise some modifications in the previous approach so that these Lie algebras appear too as the result of some `modified' construction? We first recall that initially we took as the only possibilities for each $\Ke$ the normed division algebras, but an essential step later was to extend this CD process so as to lead to the {\itshape alternative composition $*$-algebras}, which include the former together with the split forms of complex, quaternions and octonions, and the algebraic construction of the $\Ke\su_n^l$ works in the same way for them.  Is there some more freedom in the CD doubling, which could be used to still increase its extent so as to include other $*$-algebras? And can this extension, if possible, used to embrace the simple Lie algebras that up to now have not appeared in the `special unitary families' over $\K_1 \otimes \K_2$?

Up to now, only a particular antiautomorphism of $\K$ has been allowed in the game for each $\Ke$: the identity for $\Ke=\Re$, the complex conjugation for $\Ke=\Ce, \CeS$, the quaternionic conjugation for $\Ke=\He, \HeS$ and the octonionic conjugation for $\Ke=\Oe, \OeS$. These followed from the adoption of a rigid choice for the behavior under $*$ of the CD units added at each stage, and in all cases we took them to change sign under $*$. This made all the algebras to have the property of being `nicely normed' \cite{Baez:Octonions}. But these are not the only antiautomorphisms in each type. For instance, in the algebra of complex numbers, the identity map is also an antiautomorphism (usually this is seen as an automorphism, but both are the same thing for a commutative algebra). This would appear in an `enlarged' version of the CD doubling if we choose the behaviour $i^* = +i$ for the complex unit, and we can denote the result of this process as $\Cei$. Now a moment of reflection suffices to convince oneself that our construction for the Lie algebras $\Ce\su_n^l$ can be repeated without any essential change replacing everywhere $\Ce$ for $\Cei$. In this construction the new $*$ antiinvolution is the identity, and this enters as determining the `antihermitian' character of the matrices; it is clear that in this case `antihermitian' for the identity antiinvolution means antisymmetric, just as for $\Re$ and  we get for $\Cei\su_n^l$  precisely the Lie algebra $\so_n\Ce$, the orthogonal complex algebra. 

Hence this example suggests to look for the most general antiinvolutions in the algebras $\Ce$ and $\He$ (and in their split versions). 
A full list of these can be found in \cite{Porteous}. We give here a description of these results within the CD process.  

In the initial CD process $\Re \to \Ce \to \He \to \Oe$, at each stage there were no free choices: the value of the square of the adjoined new unit was equal to $-1$ and this was chosen to be pure imaginary (this is, to change sign under the $*$-conjugation in the extended algebra). Later we made a slight weakening of these conditions by allowing also new units whose square would be equal to $+1$, and this led to the `split forms' $\CeS, \HeS, \OeS$, which are algebras with some quite different properties to their `normal' siblings. That slight extension of the process was enough to provide a constructive description of the Magic Square (for which the view staying only inside the division algebras remains incomplete and quite partial). Indeed there is no compelling reason to only allow $-1$ or $+1$ as the squares of the new units. If we take any real number $\eta$ and enforce the new condition $i^2=-\eta$ for the square of the adjoined unit $i$, this slightly generalized CD doubling applied to $\Re$ would still afford (up to isomorphism) only three different systems: the complex numbers $\Ce$ when $\eta>0$, the split complex numbers $\CeS$ when $\eta<0$ and the new so-called `dual' or Study numbers for which $i^2=0$ (named after the German geometer E. Study). The dual numbers are more degenerate than its complex or split complex siblings, and if these Study numbers were allowed in the constructions described previously they would lead to algebras which are not simple, but only {\itshape contractions} of simple algebras (for instance the Euclidean algebra $\iso_n$). 

We can introduce a suitable notation for the algebras obtained by such `parame\-terized' process: all these algebras start from the reals $\Re$. The first stage adds a new unit $i_1$,  whose square is equal to a real parameter $-\eta_1$, and we may denote the algebra obtained by a symbol as $\alg{\eta_1}$, the second stage adds another new unit $i_2$ whose square is equal to another real parameter $-\eta_2$, and the algebra thus obtained may be denoted by a symbol as $\alg{\eta_2\eta_1}$, etc. We present the description of the algebras $\Ke$ obtained through one, two and three stages of the CD doubling in the following Table, where the appearance of the same algebras as soon as a single hyperbolic unit is adjoined at some stage is clear: 

\bigskip

\def\lineaAlg#1#2#3#4{{$#1$}&{#2}&{$#3$}&{$#4$}\\[2pt]} % Local Defs, working only inside group 
\def\lineaAlgA#1#2#3#4{{$#1$}&{#2}&{#3}&{$#4$}\\[2pt]}
\def\lineaAlg#1#2#3#4{{$#1$}&{#2}&{$#3$}&{$#4$}\\[2pt]} % Local Defs, working only inside group 
\def\lineaAlgA#1#2#3#4{{$#1$}&{#2}&{#3}&{$#4$}\\[2pt]}
\def\linea#1#2#3#4#5#6#7{{$#1$}&{$\ #2$}&{$#3$}&{$\ #4$}&{$\ #5$}&{$\ #6$}&{$\quad #7$}\\[2pt]}
\def\linean#1#2#3#4#5#6#7#8#9{{$#1$}&{$#6$}&{$#7$}&{$#8$}&{$#9$}\cr}

{\small{

\noindent
\begin{tabular}{llllll}
\lineaAlg{ \K }{   }{ $Description of $\K$ as Cayley-Dickson doubling of $ \Re  }{   }
\hline\hline\\[-8pt] %\SeparaBloque\SeparaBloque
\lineaAlg{ \Re }{  Reals           }{ \alg{}  }{   }
\lineaAlg{ \Ce }{  Complex         }{ \alg{+\,}  }{   }
\lineaAlg{ \CeS }{  Split complex   }{ \alg{-\,}  }{   }
\lineaAlg{ \He }{  Quaternions     }{ \alg{++\,}  }{   }
\lineaAlg{ \HeS }{  Split quaternions}{ \alg{+-\,},\  \alg{-+\,},\ \alg{--\,}  }{   }
\lineaAlg{ \Oe }{  Octonions       }{ \alg{+++\,}  }{   }
\lineaAlg{ \OeS }{  Split octonions  }{ \alg{++-\,},\ \alg{+-+\,},\ \alg{-++\,},\ \alg{+--\,},\ \alg{-+-\,},\ \alg{--+\,},\ \alg{---\,}  }{   }
\hline
\end{tabular}

}}

\bigskip

These algebras come in the ordinary CD process endowed with a particular involutory antiautomorphism, which in each case is the complex, quaternionic or octonionic conjugation. When the previous family of algebras is endowed with this antiautomorphism, the  $*$-algebras so obtained are enough to complete the general construction of the Magic Square. But as mentioned before, the Lie algebras $(\K_1\otimes\K_2)\su_n^l$ obtained this way include {\itshape most} but not {\itshape all} real forms of the simple Lie algebras; for instance those $\so_{2n}^*$ with $n$ odd (thus $2n$ is not multiple of four) do not appear in this family.
 
Motivated by this,  we can ask whether some other slight extension in the CD doubling will lead to including within a similar scheme  all real forms of simple Lie algebras. The natural freedom still available in the CD process refers to the construction of an antiinvolution in the extended algebra. The standard CD process enforces a rigid choice, and while this leads to an antiinvolution in the extended algebra, this does not give the most general antiinvolution possible for the algebra obtained at each stage. Of course, if we adopt another antiinvolution, we will have to accordingly depart from the  properties we have discussed in the standard CD doubling, so neither of these properties should be taken for granted when we allow further possibilities for $\Ke$; recall also that a $*$ at each stage is required to define the product in the next stage. Hence our description here is only heuristic. 

The next Table describes the possible choices of antiinvolutions which exists for $\Ke = \Re, \Ce, \CeS, \He,  \HeS$. 
\bigskip

{\small{
 \begin{tabular}{llllll}
\lineaAlgA{ \K }{   }{ Antiinvolution  }{   }
\hline\hline\\[-8pt] %\SeparaBloque\SeparaBloque
\lineaAlgA{ \Re  }{ Reals            }{ Identity       }{   }
\lineaAlgA{ \Ce }{ Complex          }{ Complex conjugation  }{   }
\lineaAlgA{ \CeS }{ Split complex    }{ Complex conjugation  }{   }
\lineaAlgA{ \Cei }{ Complex          }{ Identity        }{   }
\lineaAlgA{ \CeSi}{ Split complex    }{ Identity        }{   }
\lineaAlgA{ \He }{ Quaternions      }{ Quaternionic conjugation  }{   }
\lineaAlgA{ \HeS }{ Split quaternions }{ Quaternionic conjugation }{   }
\lineaAlgA{ \Her }{ Quaternions      }{ Quaternionic reversion  }{   }
\lineaAlgA{ \HeSr }{ Split quaternions }{ Quaternionic elliptic reversion  }{   }
\lineaAlgA{ \HeSh }{ Split quaternions }{ Quaternionic hyperbolic reversion   }{   }
%\lineaAlgA{ \Oc }{ Octonions        }{ Octonionic Conjugation  }{   }
%\lineaAlgA{ \Pc }{ Split Octonions   }{ Octonionic Conjugation  }{   }
\hline
\end{tabular}

}}

\bigskip
Here the emphasis should be placed in the fact that the complex or the quaternionic conjugation are not the unique antiinvolutions in these algebras. A good notation device is to append an oversymbol characterizing the involution to the usual symbol name of the algebra. When endowed with ordinary conjugation, the standard complex or quaternions will be denoted by the usual letter $\Ce, \He$, while the remaining possibilities will use some extra oversymbol. 

In particular, for complex numbers (or the split complex) with the identity antiinvolution the notations $\Cei, \CeSi$ will stand for the corresponding $*$-algebras; `antihermitian' relative to this antiinvolution will mean {\itshape complex antisymmetric}. For the quaternions, which are not commutative, the identity is not an antiinvolution, but further to the quaternionic conjugation, which changes sign to the three units, there are another antiinvolutions, called the {\itshape quaternionic reversion}, which changes the sign to just one of the three units (see \cite{Porteous}). When endowed with this antiinvolution,  quaternions will be denoted $\Her$ 
and split quaternions give rise to two possibilities, denoted $\HeSr, \HeSh$ according as to  which type of unit is reversed.  

The new $*$-algebras $\Cei, \CeSi, \Her, \HeSr, \HeSh$ are not nicely normed, and then the sums $z+z^*$ will be not real in general, so for these algebras most properties wich are taken  usually for granted for $\Ce, \He$ do not hold in the same form. In particular, as the identity map is not an antiautomorphism for quaternions, there is no strictly speaking  an `orthogonal' quaternionic algebra $\so_n\He$, and the `closest' quaternionic relative to the complex $\Cei\su_n^l\equiv \so_n\Ce$ is the algebra $\Her\su_n^l$,  which precisely gives the expected `special unitary interpretation' for the Lie algebra $\sodnstar$. This interpretation applies  in all cases, whether $n$ is even or odd (notice that when $n$ is even, this algebra had also another realization over the product of quaternions and split quaternions).  

%_________________________________________________________

\section{From the $(\K_1\myot\,\K_2) \mathfrak{sa}_{\k_1\k_2\cdots\k_{n-1}}(n)$ language  to the conventional Lie algebra naming}

The Magic Squares can be seen as the `direct part' of a dictionary translating from the $(\K_1\,\myot\,\K_2) \mathfrak{sa}_{\k_1\k_2\cdots\k_{n-1}}(n)$ language to the standard Lie algebra notation. We end by including the `inverse' dictionary, which lists the real forms of simple Lie algebras and for each entry identifies their possible $(\K_1\,\myot\,\K_2) \mathfrak{sa}_{\k_1\k_2\cdots\k_{n-1}}(n)$ realizations. As far as we know, this list has not appeared in the previous literature. 

All results will be displayed in a tabular form. We first comment on the notational conventions which allow to present a large amount of information in such a compressed form. 

First, either $\K_1$ and $\K_2$ will be an algebra taken among $\Re, \Ce, \CeS, \He, \HeS$, endowed with an antiinvolution, which usually (but not always) will be the ordinary complex, quaternionic conjugation. We take into consideration all possible  choices of a $*$-antiinvolution in the algebras $\Re, \Ce, \CeS, \He, \HeS$, so allowing the different choices for $*$ will correspond to $\Re, \Ce, \CeS, \Cei, \CeSi, \He, \HeS, \Her, \HeSr, \HeSh$, as listed in the previous section. For these non-octonionic $\Ke_1, \Ke_2$, $n$ can take any value $n\geq 2$. 

The numbers 
${\k_1, \k_2, \dots, \k_{n-1}}$ are a set of real CK labels, which should be non-zero if the algebra $(\K_1\otimes\K_2) \su_{\k_1\k_2\cdots\k_{n-1}\!}(n)$ is to be simple; there is the possibility of getting isomorphic Lie algebras for different choices of the sequence ${\k_1, \k_2, \dots, \k_{n-1}}$. To answer precisely the question of whether this is the case or not, first consider the $n\choose2$ members of the full set of {\itshape two-index CK} labels $\k_{ij}$ defined in (\ref{Metricnl}) as $\k_{ij}:= \k_{i+1}\k_{i+2}\cdots\k_{j}$ for $i, j =0, \dots n-1$, and $i<j$. In the case where all $\k_i$ are different from zero, all the $\k_{ij}$ are also different from zero. If all $\k_i$ are positive, then all the $\k_{ij}$ are also positive. But if there are some negative $\k_i$, the set of $n\choose2$  values $\k_{ij}$ will include positive as well as negative values, and the number of $\k_{ij}$ which are negative cannot be arbitrary; an easy checking tells that the number of {\it negative} $\k_{ij}$ {\it must necessarily be} of the form $l(n-l)$, with $l\leq n-l$. The number $l$ is an integer in the range between $0$ and $[n/2]$ ($[n/2]$ denotes the integer part of $n/2$). 

It turns out that two Lie algebras in the family  $(\K_1\otimes\K_2 )\su_{\k_1\k_2\cdots\k_{n-1}}(n)$ with different lists of CK labels are only isomorphic when the indices $l$ determined as above are the same (indeed, the number $l$ is precisely the {\it negative inertia index}). In the non-generic cases where at least a $\k_i$ vanishes, then some of the {\itshape two-index CK} labels $\k_{ij}$ among its $n\choose2$ members will also vanish, and there are more possibilities, into which we do not enter here (see a rather complete discussion in \cite{AHPS:CohomSO, HPS:CohomSU, HS:CohomSP}). 

We shall use some shorthands to describe the basic lists of labels $\k_1, \k_2, \dots, \k_{n-1}$ associated to any  
$$
(\K_1\otimes\K_2) \su_n^l \equiv (\K_1\myot\K_2 )\su_{\k_1 \k_2 \cdots \k_{n-1}}{(n)}.
$$
 Here a list subscripted to $\su$ will always refer to the basic list of CK labels; when the list is explicitly given $n$ appears within parentheses and $l$ ---which is univocally determined by the list--- is not displayed:

\noindent\hangindent=24pt\hangafter=1 $\plus$ means that all the $\k_i$ in the basic list are positive, and hence reducible to $+1$ by scale changes of the basic coordinates (which changes each $\k_i$ by a factor which is necessarily positive). Then the 
$n\choose2$ {\itshape two-index CK} labels $\k_{ij}$ are also all positive (and reducible by scaling to $+1$) and therefore in this case, the inertia index is equal to zero, $l=0$. 

\noindent\hangindent=24pt\hangafter=1 $\minus$ means that all $\k_i$ are negative (and hence reducible to $-1$). In this case there are positive and negative {\itshape two-index CK} labels $\k_{ij}$, and it is clear that there are in all $[n/2]$ negative values in the list $(1, \k_{12}, \k_{13}, \dots \k_{1n-1})$ (see the metric matrix \myref{Metricnl}), so that in this case $l=[n/2]$. 

\noindent\hangindent=24pt $\any$ refers to the {\itshape generic case} where all $\k_i$ are either positive or negative, but no zero values are allowed. Hence all $\k_i$ are thus reducible to either $1$ or $-1$.  

\noindent\hangindent=24pt $\indexl$ refers to any particular instance of  $\any$ where the sequence of $n$ values $(1, \k_{12}, \k_{13}, \dots, \k_{1n-1})$ which appear as the diagonal elements in the metric, has index precisey equal to $l$, this is, there are precisely $l$ negative terms. An easy checking says that this is equivalent to saying that the sequence of $\k_i$ contains no zero {\itshape and}  the number of negative terms in the {\itshape full set} of $\k_{ij}$ is $l(n-l)$.

Notice that $\plus$ or $\minus$ refer to lists for the $\k_i$ which are {\itshape completely specified}, (this is, up to the scaling factors which should reduce all the $\k_i$ to $\pm 1$)  while 
$\any$ or  $\indexl$ refer to lists which are only partially specified, with several concrete possibilities for them. In particular, we note the following relations: 

\begin{itemize}

\item 
$\plus \equiv {\{\!\!\{\n;0\}\!\!\}}$.

\item 
$\minus$ is a member appearing as one of the lists in ${\{\!\!\{\n;[n/2]\}\!\!\}}$ though for $n>2$ the shorthand ${\{\!\!\{\n;[n/2]\}\!\!\}}$  refers to several lists, with $\minus$ being just one of them. 

\item 
All the lists in $\indexl$ (for any $l$) as well as the lists $\plus$ and $\minus$ are members of $\any$.
\end{itemize}

These notational shorthands allow a clear statement on the identification of  $(\K_1\otimes\K_2)\su_{\k_1 \k_2 \cdots \k_{n-1}}{(n)}$. All throughout the tables it would be clear that in some cases the Lie algebra corresponding to a given sequence of the form $\any$ (for fixed $\K_1, \K_2$ and $n$) will be (up to isomorphism) completely independent of the particular set of (non-zero) CK labels (for instance, $ \CeS{\su}_{\any}{(n)} \equiv \sl_n\Re$, for any choice of non-zero CK labels), while in other cases the Lie algebra corresponding to a given sequence of the form $\any$ (for fixed $\K_1, \K_2$ and $n$) will (up to isomorphism) depend on the particular set of CK labels through the inertia index $l$ (as in the orthogonal case $\so_{\indexl}{(n)} \equiv \so_n^{\esig l\esig}$, where the ellipsis in the inertia index tries to remind that there is a full family of different Lie algebras in the set $\so_{\indexl}{(n)}$.

The following {\em five} tables contain a very large amount of information, and as far as we know these tables, which are a kind of reversion (and enlargement) of the information contained in the Magic Squares, have not appeared before in the literature. 

The disposition tries to be as conventional as possible, but we have to fit several peculiarities. We first deal with the classical simple Lie algebras, and the information is splitted in the {three} Tables I, II and III, each including the simple Lie algebras in the  Cartan series B \& D, A and C, respectively. Next we deal with  exceptional simple Lie algebras (all of which have some octonionic realization as a $(\K_1\otimes\K_2)\su_{\k_1 \k_2}{(3)}$, given in Table V, or as a $(\K_1\otimes\K_2)\su{(1)}$), together with some orthogonal algebras in the Cartan series B \& D which allow an exceptional octonionic realization as $(\K_1\otimes\K_2)\u_{\k_1}{(2)}$ shown in Table IV. 

In the first column, the entries are the simple Lie algebras. For classical Lie algebras we use the `dimension/inertia index' notation when pertinent, with the inertia index placed as a superscript. For the exceptional algebras, we use instead the usual notation `rank/Killing Cartan signature', and the CK signature is placed also as a superscript but enclosed in parentheses to minimize the possibility of confusions with the inertia index or any other misunderstandings. 

Only for completeness, we have also included some realizations for which either $\K_1$ or $\K_2$ or both are $\Re, \Ce, \CeS, \He, \HeS$ endowed with some antiinvolution which is not the ordinary conjugation, as explained before. 

To keep the tables within reasonable space extent, the $\K_1\otimes\K_2$ heading of columns, given simply as $\Re,\Ce,\He,\Ce\myot\Ce,\Ce\myot\He,\He\myot\He$, cover under each heading the normal and split versions  as well as all the possible 
antiinvolutions in each algebra. Thus all realizations involving say $\Ce$, $\CeS$, $\Cei$ or $\CeSi$ as $\K_1$ (and an invisible $\Re$ as $\K_2$) appear under the column heading $\Ce$. The first example in Table I is the entry $\mysa{\HeSr}{\indexl}{\n}$ appearing in the line of $\sodndl$ under the heading $\He$.   

A given Lie algebra might have several realizations as $(\K_1\otimes\K_2) \su_{\k_1 \k_2\cdots \k_{n-1}\!}{(n)}$ and these are placed either in the same line when their $\K_1\otimes\K_2$ headings are different or spanning more than one line in the same column when there are several realizations with the same   $\K_1\otimes\K_2$:    the Lie algebra $\socncl$ is the first appearance of both instances in the Tables. 

As said before, the inertia index appears as a superscript in the usual Lie algebra name. When a whole family of inertia indices are involved, this is indicated by enclosing them within an ellipsis; this way $\sonl$ might refer to either member in the set of algebras $\so_n, \,\so_n^1,\, \so_n^2, \dots$; in some cases to add some specific emphasis, the physicist standard notation $\so(p,q)$ is also indicated, with the usual convention $p>q$; the relation among both notations is $p+q=n$ and $q=l$. Quite similar notations apply for the complex unitary algebras $\su$ and the unitary quaternionic ones $\sq$, which have inertia indices as well as far as the real forms are concerned. 

Again for the completeness sake, we have included here some particular orthogonal classical Lie algebras which further to their `classical realizations' as real orthogonal algebras have another octonionic realizations as well. All these cases appear in the `unitary' (not the `special unitary') octonionic family which as we mentioned exists only for $n=2$. In this connection, a last remark to be taken into account is that a given Lie algebra might appear more than once in the full set of lists. For instance, a particular algebra like $\so_{16}^8$ appears overall in five lines in the `classical part' shown in    Table I and in one line in the `octonionic part' shown in Table IV. In addition to the `normal' entry over the reals in the first line of the Table  I for the classical series B and D, the $\so_{2n}^{2l}$ algebras with even dimension and even index have another realization over the reversed quaternions (listed in the second line), the $\so_{4n}^{4l}$ algebras with dimension and index multiple of four have still another two realizations over tensor products of quaternions and reversed quaternions (listed in the third line), and further to this the split forms $\so_{2n}^n$ and $\so_{4n}^{2n}$ have several more realizations (listed in lines four and five of the `classical' Table I). And furthermore, there is still another possible realization of $\so_{16}^8$ involving octonions, listed in the last line of the octonionic part covered by Table IV. 

%%%%%%%%%%%%%%%%%____________________________%%%%%%%%%%%%%%%%%%%%
\newpage
\leftskip-15pt \rightskip-15pt

\enlargethispage{8\baselineskip}

%\small%\footnotesize

\noindent{Table I. $(\K_1\myot\K_2) \su_{\k_1\cdots\k_{n\!-\!1}\!}(n)$ realizations of Lie algebras in classical Cartan series B\&{}D}

\noindent\begin{tabular}{lllllll}
\SeparaBloque
\linea{  }{ \ \Re }{ \Ce }{\ \ \ \ \He }{ \Ce\myot\Ce\!\! }{\ \ \ \ \ \Ce\myot\He }{\ \  \He\myot\He}\\[-18pt]
\SeparaBloque
\linea{ \sonl     }{ \mysa{\Re}{\indexl}{\n}          }{  }{  }{  }{  }{ }
\linea{ \sodndl   }{ \mysa{\Re}{\indexdl}{\dn}        }{  }{ \mysa{\HeSr}{\indexl}{\n}\hskip-20pt       }{}{}{}
\linea{ \socncl   }{ \mysa{\Re}{\indexcl}{\cn}        }{  }{ \mysa{\HeSr}{\indexdl}{\dn}\hskip-20pt   }{  }{  }{ \vtop{{ \hbox{$\mysa{\He\otimesc\He}{\indexl}{\n} $}} {\hbox{$\mysa{\HeSr\otimesc\HeSr}{\indexl}{\n} $}} } }
\linea{ \sodnn    }{ \mysa{\Re}{\indexN}{\dn}          }{  }{ \mysa{\HeSh}{\any}{\n}\hskip-20pt       }{  }{  }{  }
\linea{ \socndn   }{ \mysa{\Re}{\indexdN}{\cn}        }{  }{ \mysa{\HeSh}{\any}{\dn}\hskip-20pt      }{  }{  }{ \vtop{{ \hbox{$\mysa{\HeS\otimesc\HeS}{\any}{\n} $}}{ \hbox{$\mysa{\Her\otimesc\Her}{\any}{\n} $}}{ \hbox{$\mysa{\HeSr\otimesc\HeSh}{\any}{\n} $}}{ \hbox{$\mysa{\HeSh\otimesc\HeSh}{\any}{\n} $}} } }
\linea{ \sodnstar }{                            }{  }{ \mysa{\Her}{\any}{\n}\hskip-20pt       }{  }{  }{  }
\linea{ \socnstar }{                            }{  }{ \mysa{\Her}{\any}{\dn}\hskip-20pt     }{  }{  }{ \vtop{{ \hbox{$\mysa{\He\otimesc\HeS}{\any}{\n} $}}{ \hbox{$\mysa{\Her\otimesc\HeSr}{\any}{\n} $}}{ \hbox{$\mysa{\Her\otimesc\HeSh}{\any}{\n}  $}} } }
\linea{ \sonC     }{ }{  \hskip-20pt\mysa{\Cei}{\any}{\n}\hskip-20pt          }{                        }{  }{  }{  }
\linea{ \sodnC    }{ }{ \hskip-20pt\mysa{\Cei}{\any}{\dn}\hskip-20pt          }{  }{  }{ \hskip-0pt\vtop{{ \hbox{$\mysa{\Cei\otimesc\Her}{\any}{\n}$}}{ \hbox{$\mysa{\Cei\otimesc\HeSr}{\any}{\n} $}}{ \hbox{$\mysa{\Cei\otimesc\HeSh}{\any}{\n} $}} }\hskip-30pt  }{  }
\SeparaBloque
\end{tabular}

%%%%%%%%%%%%%%%%%____________________________%%%%%%%%%%%%%%%%%%%%
\bigskip

\noindent  {Table II. $(\K_1\myot\K_2) \su_{\k_1\cdots\k_{n-1}}(n)$ realizations of classical Lie algebras in Cartan series A}

	\begin{tabular}{llllll l}
	\SeparaBloque
	\linea{  }{ \ \Re\ \ \  }{ \ \ \ \Ce }{ \ \He\ \ \  }{\ \ \  \Ce\myot\Ce }{ \ \ \ \ \Ce\myot\He }{\He\myot\He}\\[-18pt]
	\SeparaBloque
	\linea{ \sunl     }{ }{ \mysa{\Ce}{\indexl}{\n}       }{                        }{  }{  }{  }
	\linea{ \sudndl   }{ }{ \mysa{\Ce}{\indexdl}{\dn}     }{  }{  }{ \vtop{{ \hbox{$\mysa{\Ce\otimesc\He}{\indexl}{\n} $}}{ \hbox{$\mysa{\Ce\otimesc\HeSr}{\indexl}{\n} $}} } }{  }
	\linea{ \sudnn    }{ }{ \mysa{\Ce}{\indexN}{\dn}      }{  }{  }{ \vtop{{ \hbox{$\mysa{\Ce\otimesc\HeS}{\any}{\n} $}}{ \hbox{$\mysa{\Ce\otimesc\Her}{\any}{\n} $}}{ \hbox{$\mysa{\Ce\otimesc\HeSh}{\any}{\n} $}} } }{  }
	\linea{ \sudnstar }{ }{  }{  }{  }{ \vtop{{ \hbox{$\mysa{\CeS\otimesc\He}{\any}{\n}    $}}{ \hbox{$\mysa{\CeS\otimesc\Her}{\any}{\n} $}} }   }{  }
	\linea{ \slnR     }{ }{ \mysa{\CeS}{\any}{\n}          }{  }{  }{  }{  }
	\linea{ \sldnR    }{ }{ \mysa{\CeS}{\any}{\dn}         }{  }{  }{ \vtop{{ \hbox{$\mysa{\CeS\otimesc\HeS}{\any}{\n} $}}{ \hbox{$\mysa{\CeS\otimesc\HeSr}{\any}{\n} $}}{ \hbox{$\mysa{\CeS\otimesc\HeSh}{\any}{\n} $}} } }{  }
	\\
	\linea{ \slnC     }{ }{  }{  }{ \hskip-10pt\vtop{{ \hbox{$\mysa{\Ce\otimesc\CeS}{\any}{\n}    $}}{ \hbox{$\mysa{\Ce\otimesc\Cei}{\any}{\n} $}}{ \hbox{$\mysa{\CeS\otimesc\Cei}{\any}{\n} $}} }\hskip-10pt  }{  }{  }
	\SeparaBloque
	\end{tabular}
	
\bigskip

%%%%%%%%%%%%%%%%%____________________________%%%%%%%%%%%%%%%%%%%%
\newpage

\noindent  {Table III. $(\K_1\myot\K_2) \su_{\k_1\k_2\cdots\k_{n-1}}(n)$ realizations of classical Lie algebras in Cartan series C}

	\begin{tabular}{lllllll}
	\SeparaBloque
	\linea{  }{ \ \Re\  }{ \ \Ce\  }{ \qquad\He\qquad }{\!\! \Ce\myot\Ce \!\!}{ \ \ \ \Ce\myot\He }{ \He\myot\He}\\[-18pt]
	\SeparaBloque
	\linea{ \sqnl     }{ }{ }{ \mysa{\He}{\indexl}{\n}       }{  }{  }{  }
	\linea{ \sqdndl   }{ }{ }{ \mysa{\He}{\indexdl}{\dn}     }{  }{  }{ \mysa{\He\otimesc\HeSr}{\indexl}{\n} }
	\linea{ \sqdnn    }{ }{ }{ \mysa{\He}{\indexN}{\dn}      }{  }{  }{ \vtop{{ \hbox{$\mysa{\He\otimesc\HeSh}{\any}{\n}  $}}{ \hbox{$\mysa{\HeS\otimesc\Her}{\any}{\n} $} } } }
	\linea{ \spdnR    }{ }{ }{ \mysa{\HeS}{\any}{\n}          }{  }{  }{  }
	\linea{ \spcnR    }{ }{ }{ \mysa{\HeS}{\any}{\dn}         }{  }{  }{ \vtop{{ \hbox{$\mysa{\He\otimesc\Her}{\any}{\n} $}}{ \hbox{$\mysa{\HeS\otimesc\HeSr}{\any}{\n} $}}{ \hbox{$\mysa{\HeS\otimesc\HeSh}{\any}{\n} $}} } }
	\linea{ \spdnC    }{  }{  }{  }{  }{\hskip-10pt\vtop{{ \hbox{$\mysa{\Cei\otimesc\He}{\any}{\n} $}}{ \hbox{$\mysa{\Cei\otimesc\HeS}{\any}{\n} $}} } \hskip-10pt}{  }
	\SeparaBloque
	\end{tabular}

%%%%%%%%%%%%%%%%%____________________________%%%%%%%%%%%%%%%%%%%%

\bigskip\bigskip

\noindent {Table IV. $(\K_1\otimes\K_2) \su_{\k_1\k_2}(3)$ octonionic realizations of exceptional Lie algebras in Cartan series E \& F}

	\begin{tabular}{lllllllll}
	\SeparaBloque
	\linean{  }{  }{ }{  }{ }{ \Oe }{ \Ce\myot\Oe }{ \He\myot\Oe}{ \Oe\myot\Oe}
	\SeparaBloque
	\linean{ \fivc  }{  }{  }{  }{  }{ \mysa{\Oe}{\plus}{3} }{  }{  }{  }
	\linean{ \fivxx }{  }{  }{  }{  }{ \mysa{\Oe}{\indexti}{3} }{  }{  }{  }
	\linean{ \fivs  }{  }{  }{  }{  }{ \mysa{\OeS}{\any}{3} }{  }{  }{  }
	\SeparaBloque
	\linean{ \evic    }{  }{  }{  }{  }{  }{ \hskip-10pt\mysa{\Ce\otimesc\Oe}{\plus}{3} }{  }{  }
	\linean{ \evixxvi }{  }{  }{  }{  }{  }{ \hskip-10pt\mysa{\CeS\otimesc\Oe}{\any}{3} }{  }{  }
	\linean{ \evixiv  }{  }{  }{  }{  }{  }{ \hskip-10pt\mysa{\Ce\otimesc\Oe}{\indexti}{3} }{  }{  }
	\linean{ \eviII   }{  }{  }{  }{  }{  }{ \hskip-10pt\mysa{\Ce\otimesc\OeS}{\any}{3} }{  }{  }
	\linean{ \evis    }{  }{  }{  }{  }{  }{ \hskip-10pt\mysa{\CeS\otimesc\OeS}{\any}{3} }{  }{  }
	\SeparaBloque
	\linean{ \eviic   }{  }{  }{  }{  }{  }{  }{ \mysa{\He\otimesc\Oe}{\plus}{3} }{  }
	\linean{ \eviixxv }{  }{  }{  }{  }{  }{  }{ \mysa{\HeS\otimesc\Oe}{\any}{3} }{  }
	\linean{ \eviiv   }{  }{  }{  }{  }{  }{  }{ \vtop{\hbox{$\mysa{\He\otimesc\Oe}{\indexti}{3}$ }\vskip2pt\hbox{$\mysa{\He\otimesc\OeS}{\any}{3}$ }} }{  }
	\linean{ \eviis   }{  }{  }{  }{  }{  }{  }{ \mysa{\He\otimesc\OeS}{\any}{3} }{  }
	\SeparaBloque
	\linean{ \eviiic    }{  }{  }{  }{  }{  }{  }{  }{ \mysa{\Oe\otimesc\Oe}{\plus}{3} }
	\linean{ \eviiixxiv }{  }{  }{  }{  }{  }{  }{  }{ \mysa{\Oe\otimesc\OeS}{\any}{3} }
	\linean{ \eviiis    }{  }{  }{  }{  }{  }{  }{  }{ \vtop{\hbox{$\mysa{\Oe\otimesc\Oe}{\indexti}{3}$ }\vskip2pt\hbox{$\mysa{\OeS\otimesc\OeS}{\any}{3}$ }} }{  }
	\SeparaBloque
	\end{tabular}

%%%%%%%%%%%%%%%%%____________________________%%%%%%%%%%%%%%%%%%%%
\newpage
 
\noindent {Table V. $(\K_1\otimes\K_2 )\u_{\k_1}(2)$ exceptional `unitary' octonionic realizations of some particular (orthogonal) classical Lie algebras in Cartan series B \& D and $\Oe\su(1)$ realizations of classical Lie algebras in Cartan series G}

	\begin{tabular}{lllllllll}
	\SeparaBloque
	\linean{  }{  }{ }{  }{ }{ \Oe }{ \Ce\myot\Oe }{ \He\myot\Oe}{ \Oe\myot\Oe}
	\SeparaBloque
	\linean{ \so_9  \equiv \so(9) }{  }{  }{  }{  }{ \mya{\Oe}{\plus}{2} }{  }{  }{  }
	\linean{ \so_9^1\equiv \so(8,1)  }{  }{  }{  }{  }{ \mya{\Oe}{\minus}{2} }{  }{  }{  }
	\linean{ \so_9^4\equiv \so(5,4)   }{  }{  }{  }{  }{ \mya{\OeS}{\any}{2} }{  }{  }{  }
	\SeparaBloque
	\linean{ \so_{10}  \equiv \so(10) }{  }{  }{  }{  }{  }{ \hskip-10pt\mya{\Ce\otimesc\Oe}{\plus}{2} }{  }{  }
	\linean{ \so_{10}^1\equiv \so(9,1) }{  }{  }{  }{  }{  }{ \hskip-10pt\mya{\CeS\otimesc\Oe}{\any}{2} }{  }{  }
	\linean{ \so_{10}^2\equiv \so(8,2)  }{  }{  }{  }{  }{  }{ \hskip-10pt\mya{\Ce\otimesc\Oe}{\minus}{2} }{  }{  }
	\linean{ \so_{10}^4\equiv \so(6,4)   }{  }{  }{  }{  }{  }{ \hskip-10pt\mya{\Ce\otimesc\OeS}{\any}{2} }{  }{  }
	\linean{ \so_{10}^5\equiv \so(5,5)   }{  }{  }{  }{  }{  }{ \hskip-10pt\mya{\CeS\otimesc\OeS}{\any}{2} }{  }{  }
	\SeparaBloque
	\linean{ \so_{12}  \equiv \so(12) }{  }{  }{  }{  }{  }{  }{ \hskip-10pt\mya{\He\otimesc\Oe}{\plus}{2} }{  }
	\linean{ \so_{12}^2\equiv \so(10,2)  }{  }{  }{  }{  }{  }{  }{ \hskip-10pt\mya{\HeS\otimesc\Oe}{\any}{2} }{  }
	\linean{ \so_{12}^4\equiv \so(8,4)   }{  }{  }{  }{  }{  }{  }{ \hskip-10pt\vtop{\hbox{$\mya{\He\otimesc\Oe}{\minus}{2}$ }\vskip2pt\hbox{$\mya{\He\otimesc\OeS}{\any}{2}$ }} }{  }
	\linean{ \so_{12}^6\equiv \so(6,6)   }{  }{  }{  }{  }{  }{  }{ \hskip-10pt\mya{\HeS\otimesc\OeS}{\any}{2} }{  }
	\SeparaBloque
	\linean{ \so_{16}  \equiv \so(16)   }{  }{  }{  }{  }{  }{  }{  }{ \hskip-10pt\mya{\Oe\otimesc\Oe}{\plus}{2} }
	\linean{ \so_{16}^4\equiv \so(12,4)  }{  }{  }{  }{  }{  }{  }{  }{ \hskip-10pt\mya{\Oe\otimesc\OeS}{\any}{2} }
	\linean{ \so_{16}^8\equiv \so(8,8)     }{  }{  }{  }{  }{  }{  }{  }{ \hskip-10pt\vtop{\hbox{$\mya{\Oe\otimesc\Oe}{\minus}{2}$ }\vskip2pt\hbox{$\mya{\OeS\otimesc\OeS}{\any}{2}$ }} }{  }
	\SeparaBloque
	\linean{ \giic  }{  }{  }{  }{  }{ \mysa{\Oe}{}{1} }{  }{  }{  }
	\linean{ \giis  }{  }{  }{  }{  }{ \mysa{\OeS}{}{1} }{  }{  }{  }
	\SeparaBloque
	\end{tabular}                                                                                                                                                                                                                               
\normalsize
\leftskip0pt \rightskip0pt
% >>>>>>>>>>>>>>>>>>>>>>>>>>>>>>>>>>>>>>>>>>>>>>>>>>>>>>>>>>>>>>>>>>

\section*{Acknowledgments}

The initial stages of this work were done in collaboration with F.J. Herranz \cite{TesisVulpi}. 
The paper has benefited from several pertinent comments by the referee, which are gratefully acknowledged. 

\section*{Comments on this version}

In the present version only very minor changes have been performed over the the original publication of 2013. In particular, no any change has been made in the list of references.  

% >>>>>>>>>>>>>>>>>>>>>>>>>>>>>>>>>>>>>>>>>>>>>>>>>>>>>>>>>>>>>>>>>>

\end{document}